\documentclass[aps,showpacs,prd,twocolumn]{revtex4-1}
\usepackage{eurosym}
\usepackage{amsfonts}
\usepackage{amssymb}
\usepackage{amsmath}
\usepackage{color}
\usepackage[usenames,dvipsnames]{xcolor}
\usepackage{float}
\usepackage{accents}
\usepackage{graphicx}
\usepackage{epstopdf}
\usepackage[colorlinks=true,pdfstartview=FitV,linkcolor=blue,citecolor=blue,urlcolor=blue,breaklinks=true]{hyperref}
\usepackage{ulem}

\begin{document}

\title{BPS solitons with internal structure in the gauged $O(3)$ sigma model}
\author{Rodolfo Casana}
\email{rodolfo.casana@ufma.br}\email{rodolfo.casana@gmail.com}
\author{Andr\'{e} C. Santos}
\email{andre.cs@discente.ufma.br}\email{andre$\_$cavs@hotmail.com}
\author{M. L. Dias}
\email{marcos.ld@discente.ufma.br}\email{marcosld22@gmail.com}

\affiliation{Departamento de F\'{\i}sica, Universidade Federal do Maranh\~{a}o,
65080-805, S\~{a}o Lu\'{\i}s, Maranh\~{a}o, Brazil.}

\begin{abstract}

We investigate the existence of self-dual solitons with internal structure in a gauged $O(3)$ nonlinear sigma model immersed in a dielectric medium generated by a real scalar field (dubbed the source field).  We consider rotationally symmetric configurations and applying the {Bogomol'nyi-Prasad-Sommerfield} formalism to obtain the energy lower bound and the respective {first-order differential equations (or self-dual equations).}  By solving  such a system of equations for three different dielectric media, we find the internal structure generates relevant changes in the soliton profiles when compared with the ones obtained without the presence of the dielectric medium.
\end{abstract}

\pacs{11.10.Kk, 11.10.Lm, 11.27.+d}
\maketitle

\section{Introduction}

\label{Intro}

Vortex solutions, in the field theory context, were found in the Maxwell-Higgs model by Nielsen and Olesen, relating them to the Nambu string in the strong-coupling realm \cite{Nielsen_1973}, besides in the nonrelativistic limit the Abrikosov \cite{ABRIKOSOV,Ginzburg_1955} superconducting vortices emerge naturally. In particular,  the type-II superconductors present vortex states possessing quantized magnetic flux, which was confirmed experimentally by Essmann and Tr\"{a}uble \cite{Essmann_1967_experiment}. An interesting fact about soliton solutions is that in some special situations can be obtained via a system of first-order differential equations attained employing a technique known as the Bogomol'nyi-Prasad-Sommerfield (BPS) formalism, also determining the minimum value of the system energy (the Bogomol'nyi bound) \cite{Prasad_1975, Bogo_1976}.  In connection with the Abrikosov study, the BPS limit is the interface between the two superconductor phases, such as it was studied by Bogomol'nyi in \cite{Bogo_1976} and by de Vega and Schaposnik \cite{de_Vega_1976} using an alternative technique. Furthermore, the existence of vortex solutions supporting both electric and magnetic fields also have proposed in scenarios involving the Chern-Simons action \cite{Zhang_1989, Jackiw_1990, Jackiw_1990_2, Lee_1990, Ghosh_1994}.

Vortex-like structures also emerge in the (1+2)-dimensional gauged sigma $O(3)$ model. The nonlinear sigma model has aroused the interest of lots of researchers due to its wide range of applications in condensed matter physics \cite{Polyakov:1975yp,Rajaraman:1982is, 9780852742310}. Moreover, there is a close connection between the $O(3)$ and $CP(1)$ models, such as shown in Refs. \cite{Eichenherr_1978, Golo_1978, Witten_1979}. Despite the model possesses a self-dual structure, the resulting topological solitons are scale-invariant, consequently do not represent particles in the context of Quantum Field Theory \cite{Leese_1990}.

A first intent, proposed by Schroers, breaks the scale invariance coupling minimally the sigma field to the Maxwell gauge field and introducing a potential that preserves the  self-dual structure \cite{Schroers_1995}. This way, his approach generated a new class of topological solitons with nonquantized magnetic flux. Subsequently, Ghosh \cite{Ghosh_1996} studied this new type of soliton by coupling the sigma field to the $U(1)$ Chern-Simons gauge field and, as expected, the solitons engendered - topological or nontopological - are electrically charged. For both models, the homotopy group $\pi_{2}(S^{2})$ characterizes the soliton solutions. A second approach by Mukherjee explores the breaking of the scale-invariance by adopting a self-interaction potential leading to the spontaneous breaking of the Abelian local gauge symmetry. Consequently, the topological solitons possessing quantized magnetic flux are classified through the fundamental homotopy group $\pi_{1}(S^{1})$ \cite{Mukherjee_1997, Mukherjee_1998}. These solitons also were investigated in scenarios where the sigma field is coupled nonminimally to a gauge field \cite{Cavalcante_2000} and in Lorentz-violating sigma-models  \cite{CASANA}. The BPS structure of the gauged $O(3)$ sigma model has also discussed in Ref. \cite{stepien}.

Moreover, other new vortex solutions are found promoting the extension of the  $U(1) $ symmetry, for example, in the Maxwell-Higgs model \cite{Witten_1985, Peterson_2015}. There are other interesting examples in the literature as the extended groups $U(1) \times Z_{2}$ \cite{Maxwell_Higgs_SE} and $CP(2)\times Z_{2}$ \cite{Andrade_2019}, being obtained through the introduction of a real scalar field, allowing the description of self-dual vortices in a dielectric medium. Such new objects can be of great utility in the study of metamaterials \cite{Shelby_2001, Ramakrishna_2005, Caloz_2009}.

Motivated by these discussions, we are searching for the occurrence of such first-order structures, but now in the context of the gauged {nonlinear } sigma $O(3)$ model, where its subgroup $SO(2)$ is enlarged as $SO(2)\times Z_{2}$. The corresponding extra scalar field is coupled to the gauge one by mean of a generalized dielectric function multiplying the Maxwell term.  We present our results as follows: In Sec. \ref{model}, we introduce the model, and next, by considering rotationally symmetric configurations, we implement the BPS formalism that provides the energy lower bound and the corresponding first-order equations. In Sec. \ref{solution},  we discuss three distinct scenarios by selecting the dielectric medium. After solving the BPS system of equations, we highlight the main new characteristics presented by the self-dual solitons.  Lastly, in Sec. \ref{conclusion}, we make our final comments and conclusions.

\section{The model}

\label{model}

Our starting point is a $(2+1)$-dimensional extended gauged $O(3) $ sigma model defined by the following Lagrangian density:
\begin{align}
\mathcal{L}& =-\frac{\Sigma \left( \chi \right) }{4}F_{\mu \nu }F^{\mu \nu
}+ \frac{1}{2}D_{\mu }\vec{\phi}\cdot D^{\mu }\vec{\phi}  \notag \\[0.2cm]
&\quad+\frac{1}{2}\partial _{\mu }\chi \partial ^{\mu }\chi-V(\phi _{n},\chi
),  \label{1B}
\end{align}%
where the sigma field $\vec{\phi}=( \phi _{1},\phi _{2},\phi_{3})$ is a triplet of real scalar fields whose norm is fixed to be $\vec{\phi}\cdot \vec{\phi}=1$. The tensor $F_{\mu\nu}=\partial _{\mu }A_{\nu }-\partial _{\nu }A_{\mu }$ is the electromagnetic field strength tensor of the $U(1)$ gauge field $A_{\mu}$.
The sigma and gauge fields are minimally coupling via the  covariant derivative defined as
\begin{equation}
D_{\mu }\vec{\phi}=\partial _{\mu }\vec{\phi}-A_{\mu }\hat{n}\times \vec{\phi%
},  \label{2B}
\end{equation}%
with $\hat{n}$ being an unit vector in the internal space. The self-interacting potential  $V(\phi _{n},\chi )$ is always a nonnegative function and stands for some appropriate interaction between the sigma field and the neutral scalar field $\chi$. Further, we have supposed this latter field coupled to gauge sector through a dielectric function $\Sigma(\chi)$, which is also a nonnegative real function. It is worthwhile to point out in absence of the scalar field $\chi$ we recover the standard gauged $O(3)$ sigma model studied in Ref. \cite{Mukherjee_1998}.

To investigate stationary solitons solutions we shall take $\hat{n}=(0,0,1)$ and assume henceforth the well-known hedgehog ansatz for $\vec{\phi}$,
\begin{equation}
\vec{\phi}( r,\theta) =\left(
\begin{array}{c}
\sin f(r) \cos ( N\theta ) \\[0.15cm]
\sin f(r) \sin( N\theta) \\[0.15cm]
\cos f(r)%
\end{array}%
\right) \text{,}  \label{3B}
\end{equation}%
Furthermore, for the gauge field components and the neutral
scalar field we set
\begin{equation}
A_0=A_0(r)\,,\;\; A_{i}=\epsilon _{ij}\hat{x}^{j}\frac{a(r) -N}{r}\,,\;\;
\chi=\chi(r),  \label{4B}
\end{equation}
respectively.  The quantity $N$ is a nonnull integer denoting the winding number (or topological degree). The real functions $f(r)$, $A_0(r)$, $a(r)$, and $\chi(r)$ are well-behaved satisfying appropriated boundary conditions.

We now present the stationary Euler-Lagrange equations associated with the Lagrangian density (1). The Gauss law is given by
\begin{equation}
\frac{1}{r}\left( r\Sigma A_{0}^{\prime }\right) ^{\prime }=A_{0}\sin ^{2}f%
\text{,} \label{4B0}
\end{equation}
where the prime symbol stands for the derivative with respect to the radial coordinate $r$. We observe the gauge condition $A_{0}=0$ identically satisfies the Gauss law allowing us to choose this condition for our following analysis. This way, the resulting configurations are magnetic flux carriers possessing a null electric charge.

The respective Amp\`ere law reads
\begin{equation}
\left( \Sigma B\right) ^{\prime }+\frac{a}{r}\sin ^{2}f=0,
\end{equation}
where the magnetic field, under the gauge field parametrization, becomes
\begin{equation}
B=-\frac{a^{\prime }}{r}\text{.}  \label{4B1}
\end{equation}

Under the gauge condition $A_0=0$, the field equations for the sigma and scalar field profiles, $f(r)$ and $\chi(r)$, are
\begin{equation}
\frac{1}{r}\left( rf^{\prime }\right) ^{\prime }=\frac{a^{2}}{2r^{2}} \sin(2f)+V_{f} \text{,}
\end{equation}
and
\begin{equation}
\frac{1}{r}\left( r\chi ^{\prime }\right) ^{\prime }-\frac{1}{2}B^{2}\Sigma _{\chi }=V_{\chi }\text{,}  \label{4B00}
\end{equation}
respectively. Above, we have defined $V_{f}=\partial_{f}V$, $\Sigma _{\chi} =d\Sigma /d\chi $ and $V_{\chi}=\partial _{\chi }V$.

At the origin, {the {remaining} functions} must satisfy the following boundary conditions
\begin{equation}
f\left( 0\right) =0\text{, \ \ }a\left( 0\right) =N\text{, \ \ }\chi \left(
0\right) =\chi _{0}\text{,}  \label{4B2}
\end{equation}
whereas for the asymptotic limit we require,
\begin{equation}
\lim_{r\rightarrow \infty }f(r) =\frac{\pi }{2}\text{, }\lim_{r\rightarrow
\infty }a(r) =0\text{,\ }\lim_{r\rightarrow \infty }\chi (r) =\chi _{\infty }%
\text{,}  \label{4B3}
\end{equation}%
where $\chi _{0}$ and $\chi _{\infty }$ are finite constants. Of course, the set of boundary conditions established above is consistent with the vacuum configurations of the fields and ensures the finiteness of the energy.

The corresponding energy density is
\begin{equation}
\varepsilon =\frac{1}{2}\Sigma B^{2}+\frac{1}{2}\left( f^{\prime }\right)
^{2}+\frac{a^{2}}{2r^{2}}\sin ^{2}f+\frac{1}{2}\left( \chi ^{\prime }\right)
^{2}+V.  \label{10B}
\end{equation}%
To implement the BPS procedure \cite{Bogo_1976}, we introduce two auxiliary functions $U\equiv U(f) $ and $\mathcal{W}\equiv \mathcal{W}(\chi)$, such that Eq. (\ref{10B}) can be rewritten as
\begin{align}
\varepsilon & =\frac{1}{2\Sigma }\left( \Sigma B\mp U\right) ^{2}+\frac{1}{2}%
\left( f^{\prime }\mp \frac{a}{r}\sin f\right) ^{2}  \notag \\
&\quad +\frac{1}{2}\left( \chi ^{\prime }\mp \frac{\mathcal{W}_{\chi }}{r}%
\right)^{2}+V-\frac{U^{2}}{2\Sigma }-\frac{\mathcal{W}_{\chi }^{2}}{2r^{2}}
\notag \\
&\quad \mp \frac{a^{\prime }}{r}U\mp \frac{a}{r}\left( \cos f\right)
^{\prime }\pm \frac{1}{r} \mathcal{W}^{\prime },  \label{11B}
\end{align}%
being $\mathcal{W}_{\chi }=\partial _{\chi}\mathcal{W}$. The two first terms in the third row of Eq. (\ref{11B}) can be rewritten as a total derivative by assuming the constraint $U^{\prime }=(\cos f)^{\prime }$, allowing to obtain the explicit form for $U(f)$:
\begin{equation}
U(f) =\cos f\text{,}
\end{equation}%
where, without loss of generality, we have taken the integration constant to
be zero.

Also, we choose the self-dual potential as
\begin{equation}
V\left( f,\chi \right) =\frac{U^{2}}{2\Sigma }+\frac{\mathcal{W}_{\chi }^{2}%
}{2r^{2}}=\frac{1}{2\Sigma }\cos^{2}\!f+\frac{\mathcal{W}_{\chi }^{2}}{2r^{2}%
}\text{.}  \label{11B0}
\end{equation}

The function $\mathcal{W}(\chi)$ acts as a ``superpotential" for the scalar
field $\chi$, allowing us to find solutions satisfying first-order differential equations.  It is important to mention that the presence of the neutral scalar field $\chi$ in Eq. (\ref{1B}) demands the insertion of a term, being an explicit function of the radial coordinate $r$, into the potential $V(f,\chi)$ whose finality is the full implementation of the BPS formalism. The effects of that dependence first were studied in \cite{Theocharis_2003}. Later, in Ref. \cite{Bazeia_theorem} was considered to circumvent the Derrick-Hobart theorem \cite{Hobart_1963, Derrick_1964}. Furthermore, it already has been used in different contexts, e.g., Maxwell-Higgs \cite{Maxwell_Higgs_SE, B_Research_2019}, magnetic monopoles \cite{Monopoles_SE}, and gauged $CP(2)$ \cite{Andrade_2019}.

By using the two last considerations above the energy density becomes
\begin{align}
\varepsilon & =\varepsilon_{_\text{BPS}} +\frac{1}{2\Sigma }\left( \Sigma
B\mp\cos{f}\right) ^{2}  \notag \\[0.2cm]
&\quad+\frac{1}{2}\left( f^{\prime }\mp \frac{a}{r}\sin f\right) ^{2} +\frac{%
1}{2}\left( \chi ^{\prime }\mp \frac{\mathcal{W}_{\chi }}{r}\right)^{2} ,
\label{11Bx}
\end{align}
where we have defined the BPS energy density as
\begin{equation}
\varepsilon_{_\text{BPS}}=\mp \frac{1}{r}\left(a\cos{f}-\mathcal{W}
\right)^{\prime }.  \label{zxzx}
\end{equation}
This way, from (\ref{11Bx}), we write the total energy as
follows
\begin{equation}
E= 2\pi\int_0^\infty dr\,r\,\varepsilon(r) =E_{_\text{BPS}}+\bar{E} \text{,}
\label{23Bzz}
\end{equation}%
with the BPS total energy (Bogomol'nyi bound) given by
\begin{eqnarray}
E_{_\text{BPS}}&=&\mp 2\pi\int_0^\infty dr\,r \varepsilon_{_\text{BPS}}(r),
\notag \\[0.2cm]
&=&\pm 2\pi \left( N+\Delta \mathcal{W}\right)\geq 0 \text{,}  \label{23B}
\end{eqnarray}%
where we have used the boundary conditions (\ref{4B2}) and (\ref{4B3}), and also defined $\Delta \mathcal{W}=W (\chi_{\infty})-\mathcal{W}(\chi_{0})$. The upper (lower) sign describes the self-dual solitons (antisolitons) corresponding to $N$ and $\Delta \mathcal{W}$ positive (negative) quantities. Although we have not chosen until then the explicit form for the superpotential $\mathcal{W}\left( \chi \right) $ and dielectric function $\Sigma(\chi)$, we emphasize that the BPS total energy depends only on the boundary conditions of $\mathcal{W}(\chi)$, besides, of course, the winding number $N$.

Coming back to Eq. (\ref{23Bzz}), the second term reads
\begin{eqnarray}
\bar{E} &=&2\pi\int_0^\infty dr\,r\left\{ \frac{1 }{2\Sigma}\left( \Sigma
B\mp \cos f\right) ^{2}\right.  \notag \\[0.2cm]
&&\left.\quad\quad+\frac{1}{2}\left( f^{\prime }\mp \frac{a}{r}\sin f\right)
^{2} +\frac{1}{2}\left( \chi ^{\prime }\mp \frac{\mathcal{W}_{\chi }}{r}%
\right) ^{2}\right\} \text{,}\quad\quad\quad
\end{eqnarray}%
such that the total energy (\ref{23Bzz}) has a bound $E\geq E_{_\text{BPS}}$
which  is saturated when $\bar{E}=0$, i.e., the solutions obeys the
first-order differential equations
\begin{align}
f^{\prime }& =\pm \frac{a}{r}\sin f\text{,}  \label{20B} \\[0.2cm]
B& =\pm \frac{\cos f}{\Sigma }\text{,}  \label{19B} \\[0.2cm]
\chi ^{\prime }& =\pm \frac{\mathcal{W}_{\chi }}{r}\text{.}  \label{21B}
\end{align}%
This set is called self-dual or BPS equations, where the (lower) upper sign stands for (anti) solitons with ($N<0$) $N>0$. Therefore, the BPS system ensures the energy lower bound and the stability of the corresponding field configurations. A comment not least is that indeed the Eqs. (\ref{20B})-(\ref{21B}) satisfy the set of Euler-Lagrange equations associated with Lagrangian density (\ref{1B}) as well.

Furthermore, by using the BPS equations, the BPS energy density (\ref{zxzx}) can be rewritten in the form
\begin{equation}
\varepsilon _{_{\text{BPS}}}=\varepsilon _{_\Sigma }+\varepsilon _{\chi },\label{22B}
\end{equation}
where we have defined
\begin{equation}
\varepsilon _{_\Sigma }=\Sigma B^{2}+\frac{a^{2}}{r^{2}}\sin ^{2}f\quad\text{and}\quad\varepsilon _{\chi }=\frac{\mathcal{W}_{\chi }^{2}}{r^{2}}{,}
\label{23B0}
\end{equation}%
respectively. The $\varepsilon _{_\Sigma }$ standing for the energy density associated with the new solitons while $\varepsilon _{\chi }$ is the contribution belonging to the kink $\chi$ that, according to the BPS equation (\ref{21B}), becomes independent of the other fields. This situation allows us to choose some convenient $\chi$-field configurations sourcing the dielectric medium driving the new soliton configurations.

At this stage, we observe both the dielectric function and the self-dual equation (\ref{21B}) do not include either the sigma field and the gauge field. Such a situation enables us to study interesting physical scenarios by adequately selecting the dielectric function and the superpotential. In the next section, we shall address some interesting scenarios by solving the Eq. (\ref{21B}) for a given superpotential $\mathcal{W} (\chi)$  and choosing some different dielectric functions $\Sigma (\chi)$.

\section{Some scenarios with internal structures}

\label{solution}

We shall consider some internal structure scenarios by choosing a specific form of the $\chi $ field that introduces additional nonlinearities to the original sigma model, allowing us to analyze how the shape of the original solitons is modified. For this purpose, we consider the following superpotential:
\begin{equation}
\mathcal{W}\left( \chi \right) =\alpha \chi -\frac{\alpha }{3}\chi ^{3}\text{%
,}  \label{24B}
\end{equation}%
where $\alpha $ is a positive parameter. The particular case $\alpha =1$ has  been previously approached in different contexts as global defect structures \cite{Bazeia_theorem}, skyrmion-like configurations \cite{Bazeia_2016, Bazeia_2017}, massless Dirac fermions \cite{Bazeia_2018}, magnetic monopoles \cite{Monopoles_SE} and vortices with internal structures \cite{Maxwell_Higgs_SE,Andrade_2019}. On the other hand, arbitrary values of $\alpha$ were used to discuss the solutions into a multilayered structure \cite{B_Research_2019}.

Then, by assuming the superpotential (\ref{24B}), the BPS equation (\ref{21B}) results
\begin{equation}
\chi ^{\prime }=\pm \frac{\mathcal{\alpha }}{r}\left( 1-\chi ^{2}\right)
\text{,}
\end{equation}%
which implies in the exact kink-like solution%
\begin{equation}
\chi(r) =\pm \frac{r^{2\alpha }-r_{0}^{2\alpha }}{r^{2\alpha}+r_{0}^{2\alpha
}}\text{,}  \label{25B}
\end{equation}
where $r_{0}$ is an arbitrary positive constant. Besides the solution satisfies $\chi \left( r_{0}\right) =0$, it fixes the boundary conditions for the neutral field: $\chi(0)=\chi _{0}=\mp 1$ and $\chi(\infty)=\chi _{\infty }=\pm 1$.

Under such considerations, the BPS bound for the energy (\ref{23B}) becomes%
\begin{equation}
E_{_\text{BPS}}=2\pi \left\vert N\right\vert +\frac{8}{3}\alpha \pi ,
\label{25B0}
\end{equation}%
where the second term is the contribution from the neutral scalar field. Similarly, the magnetic flux reads
\begin{equation}
\Phi =\int d^{2}xB=2\pi \left\vert N\right\vert.
\end{equation}

For our study, in the remaining of the manuscript, we only consider the soliton solutions, i.e., $N>0$.

\subsection{First scenario\label{Fs}}

We begin by setting the dielectric function to be used in our first scenario,
\begin{equation}
\Sigma \left( \chi \right) =\frac{1}{ 1-\chi ^{2}} .  \label{26B0}
\end{equation}
We note that the dielectric function diverges at the boundary values but, despite that, the BPS energy density BPS (\ref{22B}) remains finite because of the magnetic field $B(r)$ controls these singularities (as we see later).

Within this scenario, we only consider the kink solution (\ref{25B}) with $\alpha=1$ because there are not solutions acceptable physically  when $\alpha\geq2$. Thus, the dielectric function (\ref{26B0}) becomes
\begin{equation}
\Sigma \left( \chi \right)=\frac{(r^{2} +r_{0}^{2})^2}{4r^{2} r_{0}^{2}} .
\label{26B}
\end{equation}

To obtain the corresponding BPS solutions to the sigma and gauge fields, we
must solve the equations (\ref{20B}) and (\ref{19B}) by considering the
dielectric function (\ref{26B}).  Thus, the new system reads
\begin{eqnarray}
f^{\prime }& =&\frac{a}{r}\sin f\text{,}  \label{32ax} \\[0.2cm]
-\frac{a^{\prime }}{r}&=& \frac{4r^{2}r_{0}^{2}}{\left(
r^{2}+r_{0}^{2}\right) ^{2}}\cos f\text{,}  \label{33ax}
\end{eqnarray}%
where also has been used (\ref{4B1}). The system above must be solved
obeying the boundary conditions of the fields $a(r)$ and $f(r)$, namely the
Eqs. (\ref{4B2}) and (\ref{4B3}).

We now show the field behaviors in the proximity of the boundary values. Near the origin, the sigma field behaves as
\begin{eqnarray}
f(r) &\approx&f_{N}r^{N}-\frac{f_{N}r^{N+4}}{4r_{0}^{2}}+\frac{2f_{N}r^{N+6}} {9r_{0}^{4}} -\frac{(f_{N}) ^{3}r^{3N}}{12}\nonumber\\[0.2cm]
&& +\frac{\left( N^{2}+4N+12\right) (f_{N}) ^{3}r^{3N+4}}{16\left(N+2\right) ^{2}r_{0}^{2}} \nonumber\\[0.2cm]
& & +\frac{(f_{N}) ^{5}r^{5N}}{80} -\frac{(f_{N}) ^{7}r^{7N}}{448},\quad \quad\label{fN0}
\end{eqnarray}
where $f_{N}$ is a positive parameter, which can be determined numerically. For the gauge field profile, we have
\begin{eqnarray}
a(r)  &\approx&N-\frac{r^{4}}{r_{0}^{2}}+\frac{4r^{6}}{3r_{0}^{4}}-%
\frac{3r^{8}}{2r_{0}^{6}} +\frac{(f_{N}) ^{2}r^{2N+4}}{\left( N+2\right) r_{0}^{2}} \nonumber\\[0.2cm]
&& -\frac{2(f_{N}) ^{2}r^{2N+6}}{\left( N+3\right) r_{0}^{4}}   -\frac{(f_{N}) ^{4}r^{4N+4}}{8\left( N+1\right) r_{0}^{2}} .\quad \quad
\end{eqnarray}%
Both expressions above guarantee  at least the three first lowest-order terms for all values of $N$.

\begin{figure}[t]
\includegraphics[width=4.2cm]{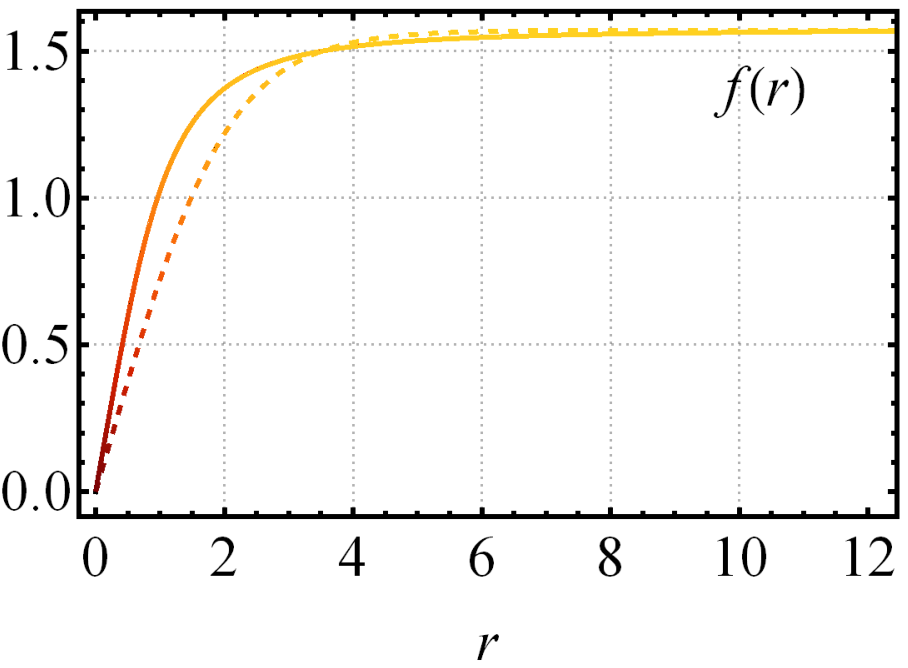}\hspace{0.14cm} %
\includegraphics[width=4.2cm]{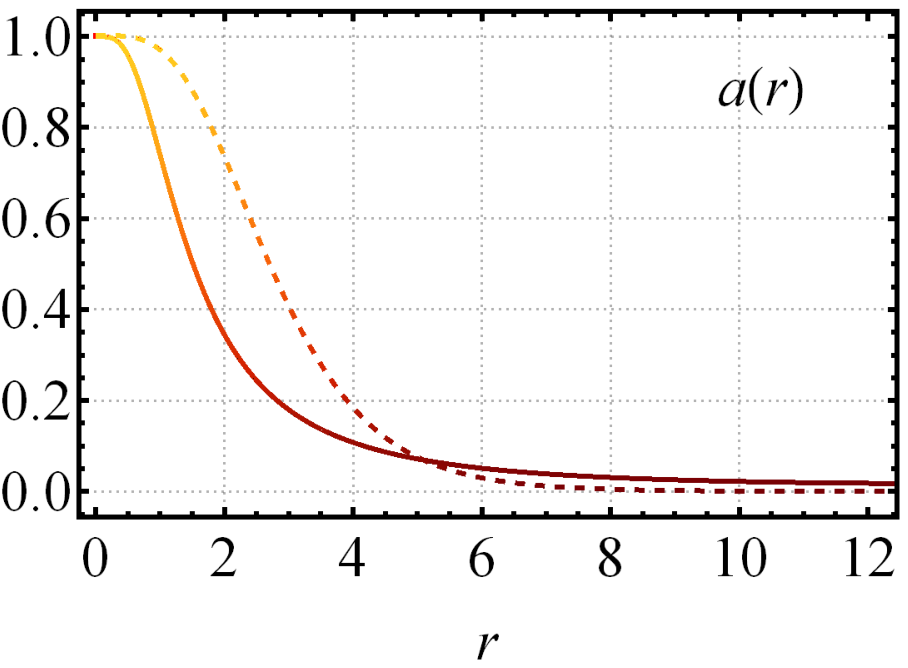} \vspace{-0.6cm}
\caption{The profiles $f(r) $ (left) and $a(r) $
(right) {for $N=1$,} $r_{0}=1$ (solid line) and $r_{0}=5$ (dashed line).}
\label{Fig01}
\end{figure}

The behavior of the fields in the asymptotic limit is given by
\begin{eqnarray}
f(r) &\approx &\frac{\pi }{2}-C_{\infty }r^{-2r_{0}}, \\[0.2cm]
a(r) &\approx &2r_{0}C_{\infty }r^{-2r_{0}},  \label{bh2}
\end{eqnarray}%
where $C_{\infty}$ is a positive constant whose value depends on the winding number. We point out the asymptotic behavior in the original gauged sigma model \cite{Mukherjee_1997} follows an exponential-law decay, which is very similar to the one shown by the Abrikosov-Nielsen-Olesen vortices \cite{ABRIKOSOV, Nielsen_1973}. However, in this case, the dielectric medium changes the asymptotic behavior of the field profiles that now follow a power-law decay.

\begin{figure}[t]
\includegraphics[width=4.2cm]{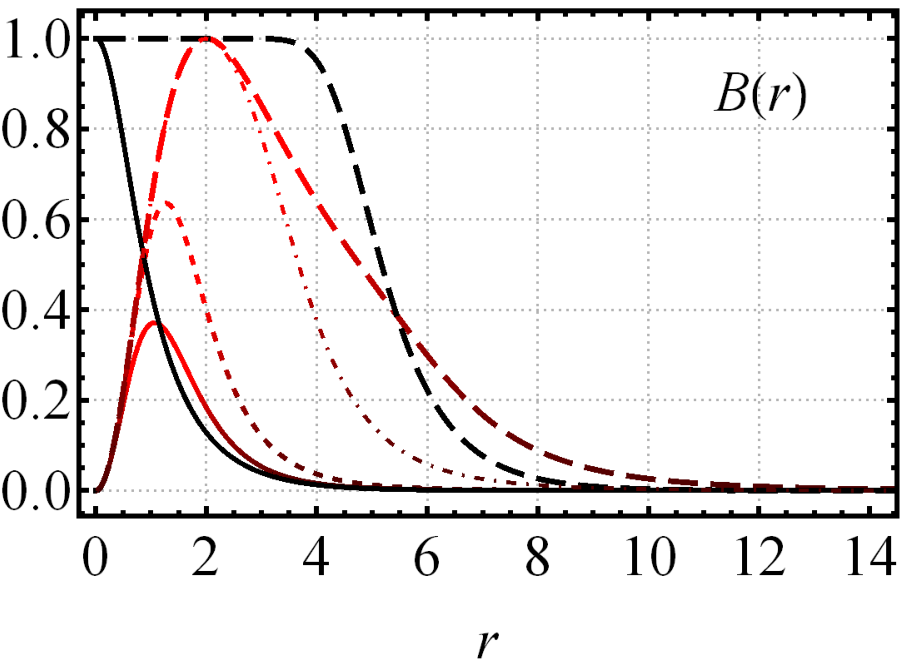}\hspace{0.15cm} %
\includegraphics[width=4.2cm]{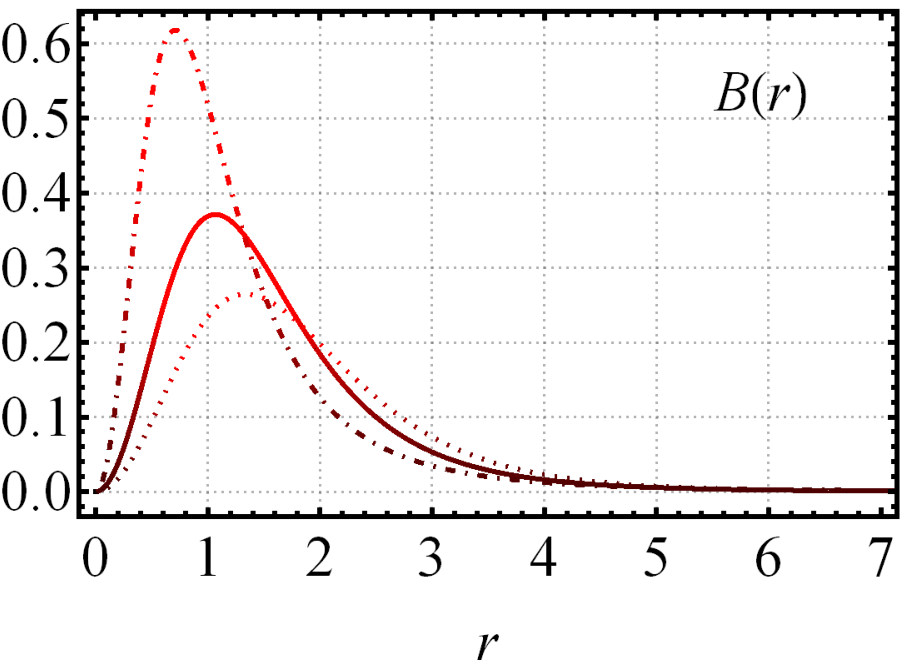}\vspace{-0.2cm}
\caption{The magnetic field profiles are depicted for some values of $N$ and $r_{0}$. There are exhibited the ones obtained both in the presence (color lines) and in the absence  of the dielectric function (black lines).  On the left, we depict for $r_{0}=2$, $N=1$ (solid line), $N=2$ (dashed line), $N=8$ (dot-dashed line) and $N=15$ (long-dashed line). On the right, for $N=1$, $r_{0}=1$ (dot-dashed line), $r_{0}=2$ (solid line) and $r_{0}=3$ (dotted line).}
	\label{Fig02}
\end{figure}

\begin{figure}[b]
\includegraphics[width=3.8cm]{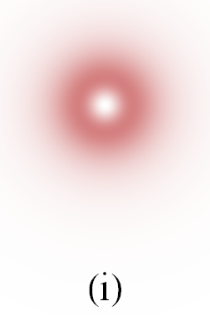}\hspace{0.cm} %
\includegraphics[width=3.8cm]{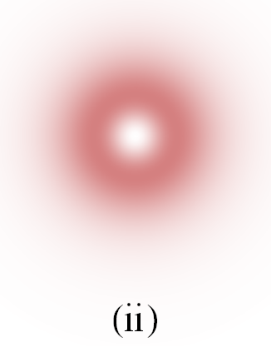}\vspace{-0.2cm}
\caption{The magnetic field depicted in the plane for $N=1$, $r_{0}=1$ (i)
and $r_{0}=2$ (ii).}
\label{Fig03}
\end{figure}

\begin{figure}[t]
\includegraphics[width=4.2cm]{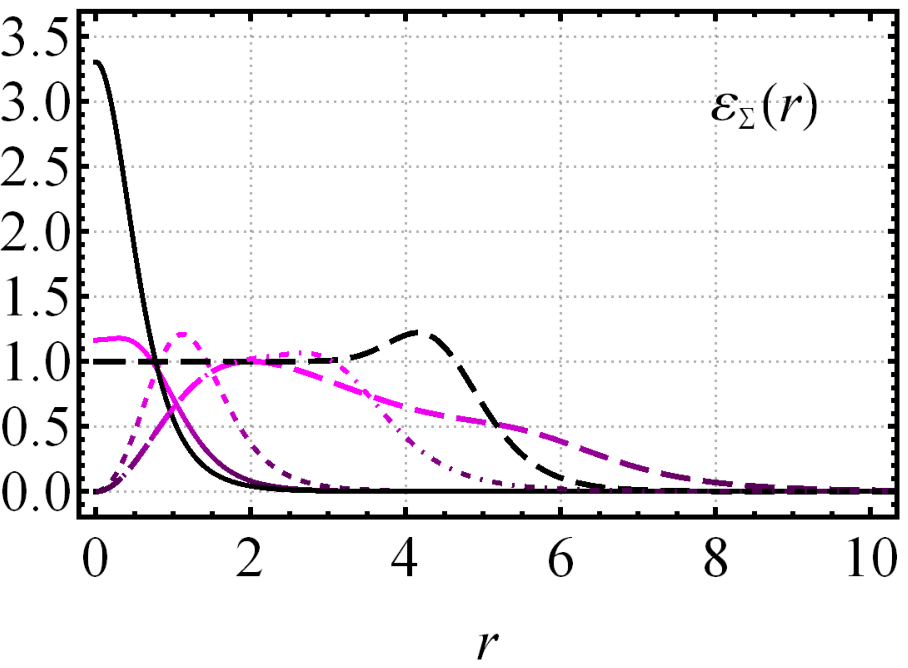}\vspace{0cm}
\includegraphics[width=4.2cm]{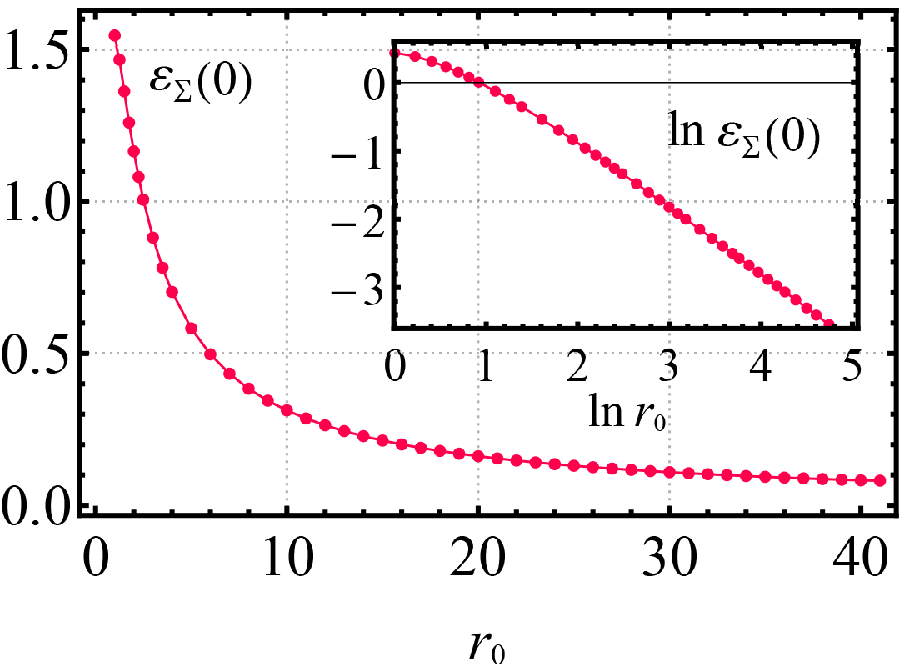} \vspace{-0.25cm}
\caption{The profiles for the  energy density $\varepsilon_{_\Sigma} (r)$.  There are exhibited the ones obtained both in the presence (color lines) and in the absence  of the dielectric function (black lines). On the left, the conventions are as in Fig. \ref{Fig02}. On the right, we depict $\varepsilon_{_\Sigma} (0) $ v.s. $r_{0}$. The insertion shows $\ln\varepsilon_{_\Sigma} (0)$ v.s. $\ln r_{0}$.}
\label{Fig04}
\end{figure}

In what follows, we present the numerical solution of the system formed by Eqs. (\ref{32ax}) and (\ref{33ax}) for some values of $N$ and $r_0$. The resulting field profiles for the gauge and sigma fields, magnetic field, and {energy density $\varepsilon_{_\Sigma} (r)$ are} shown in Figs. \ref{Fig01}--\ref{Fig04}.

Without loss of generality, we consider the field profiles $f(r)$ and $a(r)$ for the winding number {$N=1$ and} distinct values of $r_{0}$, see Fig. \ref{Fig01}. We remark that both profiles are well-behaved according to the respective boundary values,  but a new effect is observed in the gauge field profiles when compared to the ones found without the dielectric medium. Such an effect is a plateau extending from the origin whose length increases as  $r_{0}$ grows, and it directly impacts the shape of the magnetic field profile.

Figure \ref{Fig02} shows the magnetic field profiles for different values of $N$ and $r_{0}$. Differently to the case without the dielectric medium (i.e., from standard sigma model \cite{Schroers_1995}), the magnetic field is null at $r=0$ and the profiles acquire the format of rings centered at the origin. By considering  a fixed $r_0$, there is a winding number $N^{(B)}_0$ (e.g., $N^{(B)}_0=8$ in Fig. \ref{Fig02}) allowing us to distinguish the behavior of the profiles. For $N\leq N^{(B)}_0$, the maximum amplitude (located at $r=r^*\leq r_0$) increases as $N$ grows meanwhile runs far away from the origin until that {at} $N=N^{(B)}_0$ attains the maximum value equal to unity (the one in the standard sigma model) and becomes located at $r^*=r_0$.  Already for $N>N^{(B)}_0$, the maximum amplitude remains located at $r^*=r_0$ and with the same maximum value (see the left-hand side in Fig. \ref{Fig02}). On the other hand,  for a fixed $N$, the maximum of the ring decreases as  $r_0$ increases (see the right-hand side in Fig. \ref{Fig02}). Alternatively, a planar depiction provides a better view of the ringlike structure of the magnetic field, such as shown in Fig. \ref{Fig03}.

The null value at the origin of the magnetic field is corroborating by the behavior at $r=0$ given by
\begin{eqnarray}
B(r) &\approx&\frac{4r^{2}}{r_{0}^{2}}-\frac{8 r^{4}}{r_{0}^{4}}+\frac{12 r^{6}}{r_{0}^{6}} -\frac{2(f_{N}) ^{2}r^{2N+2}}{r_{0}^{2}}  \notag\\[0.2cm]
&&+\frac{4(f_{N}) ^{2}r^{2N+4}}{r_{0}^{4}} +\frac{\left(f_{N}\right) ^{4}r^{4N+2}}{2r_{0}^{2}} ,\quad\quad
\end{eqnarray}
valid at least for the three first lowest-order terms.

In this scenario, the magnetic field behaves very similarly to the one presented in the Chern-Simons-Higgs model \cite{Jackiw_1990} despite the magnetic field vanishes asymptotically following a power-law,
\begin{equation}
B(r) \approx \frac{4C_{\infty }r^2_{0}}{r^{2+2r_{0}}}.
\end{equation}%

Concerning the energy density $\varepsilon_{_\Sigma}(r)$, its behavior near  the origin (valid at least for the three first lowest-order terms) is given by
\begin{eqnarray}
\varepsilon_{_\Sigma}(r) &\approx&N^{2}(f_{N}) ^{2}r^{2N-2}+
\frac{4 r^{2}}{r_{0}^{2}}-\frac{8 r^{4}}{r_{0}^{4}}  \notag \\[0.2cm]
&& -\frac{N^{2}(f_{N}) ^{4}r^{4N-2}}{2} +\frac{3N^{2}\left(
f_{N}\right) ^{6}r^{6N-2}}{16}  \notag \\[0.2cm]
&&  -\frac{\left( N^{2}+4N+8\right) (f_{N}) ^{2}r^{2N+2}}{2r_{0}^{2}}  ,\quad\quad
\end{eqnarray}
in according to the profiles shown in Fig. \ref{Fig04}. On the left-hand side, for a fixed $r_0$ and $N=1$, the $\varepsilon_{_\Sigma}$ profiles are {nonnull} at the origin, having a lump-like format with the center slightly away from $r=0$. However, with a fixed $r_0$ and $N\geq 2$, they are  nulls at $r=0$, acquiring a ring-like format. Further, on the right of Fig. \ref{Fig04}, we observe that, for $N=1$, the values of $\varepsilon _{_\Sigma}(0) =f_{1}^{2}$ decrease as $r_{0}$ grows, and for sufficiently large values of $r_0$ decays as $r_{0}^{-1}$  (see inset). Already for the asymptotic limit, the energy density  $\varepsilon_{_\Sigma}$ follows the behavior
\begin{equation}
\varepsilon_{_\Sigma}(r) \approx \frac{8(C_{\infty })^2 r_{0}^{2}}{r^{2+4r_0}}\text{,}
\label{ebpsI}
\end{equation}%
valid for any value $N$.


\subsection{Second scenario\label{Fs2}}

In this section, we analyze the BPS configurations arising in a second dielectric medium mapped by the function
\begin{equation}
\Sigma(\chi)=\frac{1}{\chi^{2}}=\frac{\left(r^{2\alpha}+r_{0}^{2\alpha} \right) ^{2}}{\left( r^{2\alpha }-r_{0}^{2\alpha }\right) ^{2}}\text{,}
\label{27B}
\end{equation}%
where the kink $\chi (r)$ is given by Eq. (\ref{25B}). Distinctly to the
previous case, the dielectric function is finite at $r=0$ and $r\rightarrow \infty $, but possesses a divergence at $r=r_{0}$. The existence of self-dual configurations with finite energy density (\ref{22B}) is not affected by such a singularity. Indeed, this is quickly verified by analyzing the BPS equation (\ref{19B}) after substituting the Eq. (\ref{27B}),  which leads us to
\begin{equation}
B(r)=\frac{\left( r^{2\alpha }-r_{0}^{2\alpha }\right) ^{2}}{\left(
r^{2\alpha }+r_{0}^{2\alpha }\right) ^{2}}\cos f\text{,}  \label{29Bx}
\end{equation}%
showing the magnetic field is null in $r=r_{0}$, $B(r_{0})=0$, consequently,  the term $\Sigma B^{2}$ in (\ref{22B}) becomes not singular and the total BPS energy (\ref{25B0}) remains finite. Further, we here point out the vanishing of the magnetic field in this particular point will reveal the strong influence of the dielectric medium in the structure of the new soliton solutions.

The set of BPS equations describing the new solitons in this
second scenario is given by
\begin{eqnarray}
f^{\prime } &=&\frac{a}{r}\sin f\text{,}  \label{32ay} \\[0.2cm]
-\frac{a^{\prime }}{r} &=&\frac{\left( r^{2\alpha }-r_{0}^{2\alpha }\right)
^{2}}{\left( r^{2\alpha }+r_{0}^{2\alpha }\right) ^{2}}\cos f\text{.}
\label{29B}
\end{eqnarray}

For a clearer understanding of the field behaviors near the boundary values  (\ref{4B2}) and (\ref{4B3}), we need to solve the equations  (\ref{32ay}) and (\ref{29B}). This way, we first obtain the behaviors of the sigma and gauge fields near the origin,
\begin{eqnarray}
f(r) &\approx &f_{N}r^{N}-\frac{f_{N}r^{N+2}}{4}+\frac{f_{N}r^{N+4}}{32} -\frac{(f_{N}) ^{3}r^{3N}}{12}\notag \\[0.2cm]
&& +\frac{(N^{2}+2N+3)(f_{N}) ^{3}r^{3N+2}}{16(N+1)^{2}}  +\frac{(f_{N}) ^{5}r^{5N}}{80}\notag \\[0.2cm]
&& +\frac{f_{N}r^{N+2\alpha +2}}{(\alpha +1)^{2}r_{0}^{2\alpha }} ,\quad
\end{eqnarray}%
and
\begin{eqnarray}
a(r) &\approx&N-\frac{r^{2}}{2}+\frac{(f_{N}) ^{2}r^{2N+2}}{4(N+1) } -\frac{(f_{N}) ^{2}r^{2N+4}}{8 ( N+2 ) }  \notag\\[0.2cm]
&& +\frac{2r^{2\alpha +2}}{\left( \alpha +1\right) r_{0}^{2\alpha }} -\frac{4r^{4\alpha +2}}{\left( 2\alpha +1\right) r_{0}^{4\alpha }} +\ldots\notag\\[0.2cm]
&& -\frac{(f_{N}) ^{2}r^{2N+2\alpha +2}}{\left( N+\alpha +1\right) r_{0}^{2\alpha }} -\frac{(f_{N}) ^{4}r^{4N+2}}{16\left( 2N+1\right) } ,\quad\quad
\end{eqnarray}
respectively, where the quantity $f_{N}$ stands for a positive constant. These expressions guarantee at least the three first lowest-order terms of the behavior of the field profiles.

\begin{figure}[t]
\includegraphics[width=4.2cm]{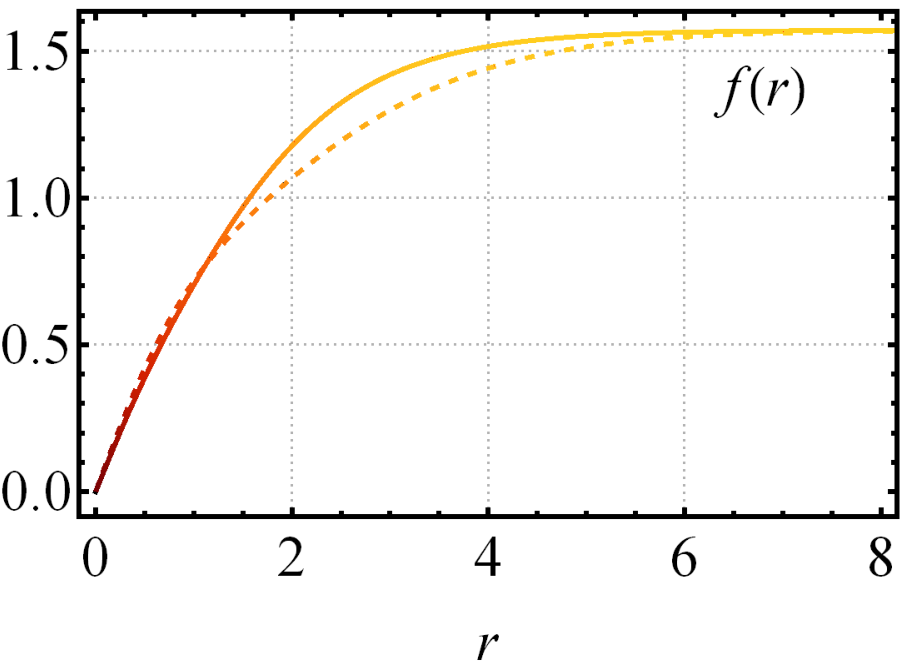}\hspace{0.14cm}
\includegraphics[width=4.2cm]{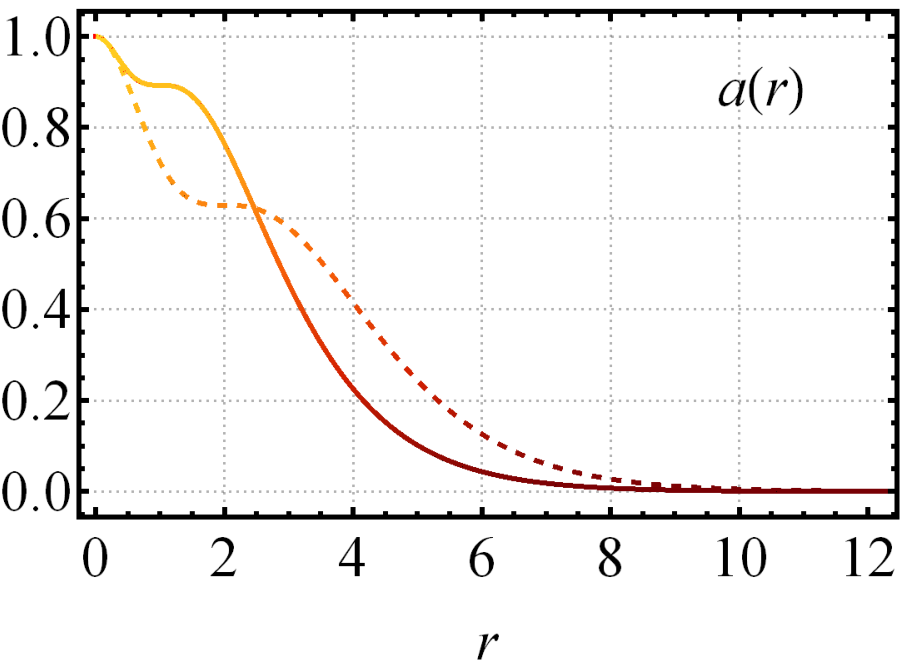} \vspace{-0.6cm}
\caption{The profiles $f(r)$ (left) and $a(r)$
(right) for $N=1$ and $\alpha=1$: $r_{0}=1$ (solid line) and $r_{0}=2 $
(dashed line).}
\label{Fig05}
\end{figure}

\begin{figure}[b]
	\includegraphics[width=4.2cm]{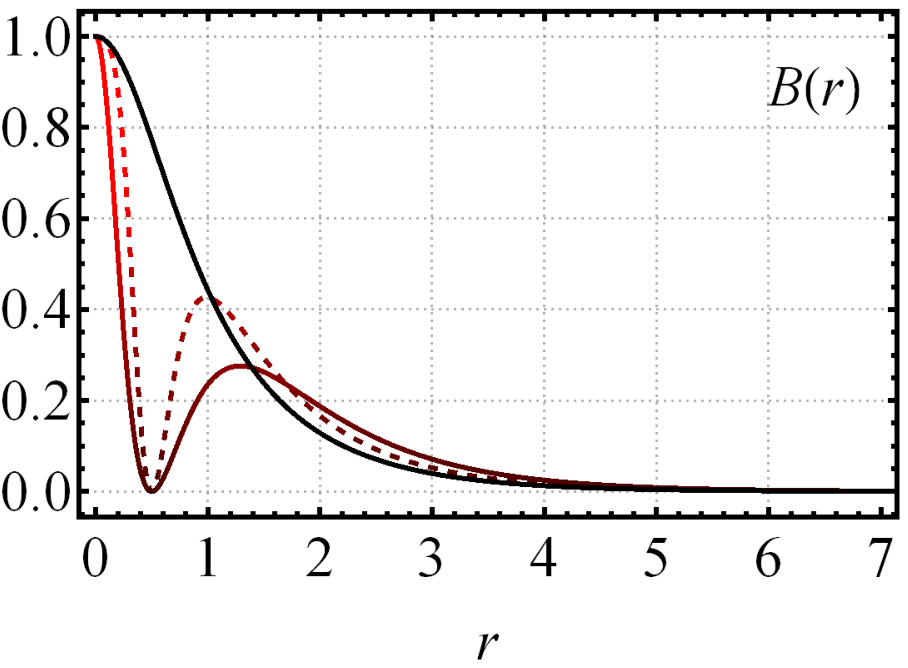}\hspace{0.14cm} %
	\includegraphics[width=4.2cm]{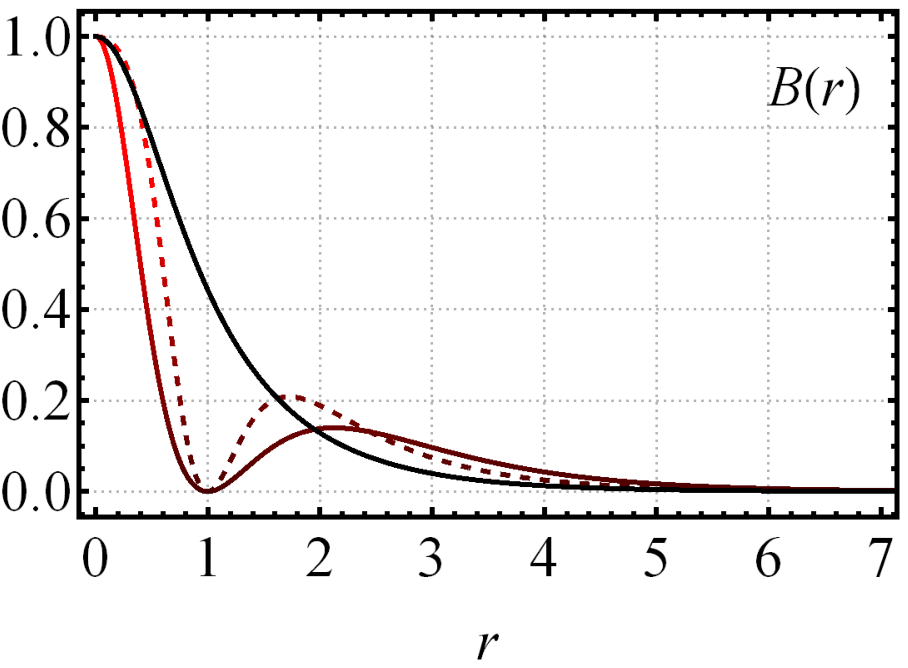} \vspace{-0.6cm}
	\caption{The magnetic field profiles for the sigma model for both in the presence (color line) and absence (black line) of the dielectric function (\ref{27B}) with fixed value $N=1$. For the case with a dielectric medium, we depict $r_{0}=0.5$ (left) and $r_{0}=1 $
		(right) with $\alpha =1$ (solid line) and $\alpha =2$ (dashed line). }
	\label{Fig06}
\end{figure}

Meanwhile, for $r\rightarrow\infty$ and all values of $N$ and $\alpha$, the
behavior of the field profiles obeys
\begin{eqnarray}
f(r) &\approx &\frac{\pi}{2}-\mathcal{C}_{\infty }r^{-1/2}e^{-r}\text{,} \label{case2inf_1}\\[0.3cm]
a(r) &\approx &\mathcal{C}_{\infty }r^{1/2}e^{-r}\text{,} \label{case2inf_2}
\end{eqnarray}
being $\mathcal{C}_{\infty }$ a positive constant. Interestingly, in this case, the asymptotic behavior of the field profiles is very similar to the one shown by the Abrikosov-Nielsen-Olesen vortices \cite{ABRIKOSOV,Nielsen_1973}, i.e., the dielectric medium does not change the asymptotic behavior which remains the same of the original gauged sigma model \cite{Mukherjee_1997}.

In what follows, we show the numerical solutions of the system formed by Eqs. (\ref{32ay}) and (\ref{29B}). For $N=1$, $\alpha =1$, and different values of $r_{0}$, the gauge and sigma field profiles are depicted in Fig. \ref{Fig05} (the value $\alpha=1$ is enough to investigate the main features of the field profiles in the current scenario). The novelty arises in the gauge field profiles that exhibit a quirky behavior: the emergence of a plateau effect around $r_{0}$ implying in important modifications in the magnetic field behavior, as we will notice afterward.

\begin{figure}[t]
	\includegraphics[width=3.8cm]{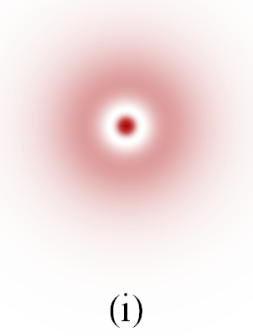}\hspace{0.cm} %
	\includegraphics[width=3.8cm]{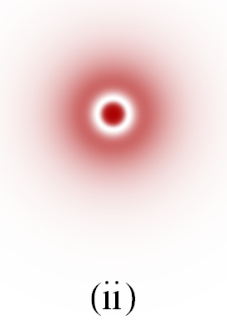}\hspace{0.cm} %
	\includegraphics[width=3.8cm]{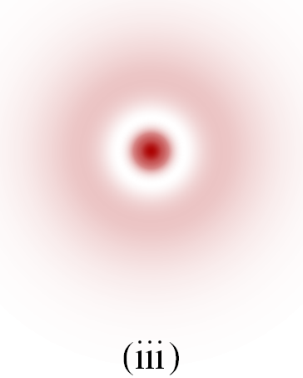}\vspace{0.cm} %
	\includegraphics[width=3.8cm]{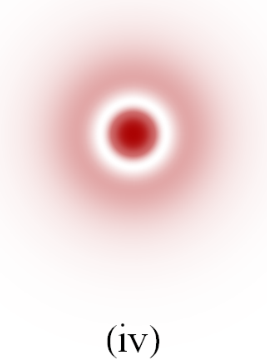}\vspace{0.cm}
	\caption{The magnetic field depicted in the plane for $r_{0}=0.5$, $%
		\alpha=1$ (i) and $\alpha=2$ (ii), and $r_{0}=1$, $\alpha=1$
		(iii) and $\alpha=2$ (iv).}
	\label{Fig07}
\end{figure}

Figure \ref{Fig06} depicts the magnetic field profiles for distinct values of the parameters $r_{0}$ and $\alpha$ and the corresponding ones of the standard sigma model. Unlike the previous scenario, we now see that the dielectric medium causes a second maximum located at $r^*>r_0$, whose amplitude is lower than the first one located at the origin. The absolute maximum at $r=0$ is verified explicitly by the magnetic field behavior, which reads as,
\begin{eqnarray}
B(r) &\approx&1-\frac{(f_{N}) ^{2}r^{2N}}{2}+\frac{(f_{N})^{2}r^{2N+2}}{4} -\frac{4r^{2\alpha }}{r_{0}^{2\alpha }}\notag \\[0.2cm]
&&+\frac{8r^{4\alpha }}{r_{0}^{4\alpha }} +\frac{2(f_{N}) ^{2} r^{2N+2\alpha}}{r_{0}^{2\alpha }}  +\frac{(f_{N}) ^{4}r^{4N}}{8} ,\label{Bii}
\end{eqnarray}
ensuring at least the two first lowest-order terms. The magnetic field profiles appear similar to the Nielsen-Olesen configurations for $0\leq r \leq r_{0}$ while in the region $r>r_{0} $ resembles the ones in the Chern-Simons-Higgs model. In the last region, the corresponding behavior for $r\rightarrow\infty$ reads
\begin{equation}
B(r) \approx  \mathcal{C}_{\infty} r^{-1/2}e^{-r},\label{BiiInfy}
\end{equation}%
being similar to the one presented in the absence of the dielectric medium. Alternatively, Fig. \ref{Fig07} provides an overview of the effects induced in the magnetic field profiles by the dielectric function (\ref{27B}) via the parameters $r_{0}$ and $\alpha$. We note that $r_{0}$ controls the internal size of the structures, while $\alpha$ controls the core size and the maximum of the external ring surrounding it, such that they increase as $\alpha$ grows.

\begin{figure}[t]
	\includegraphics[width=4.2cm]{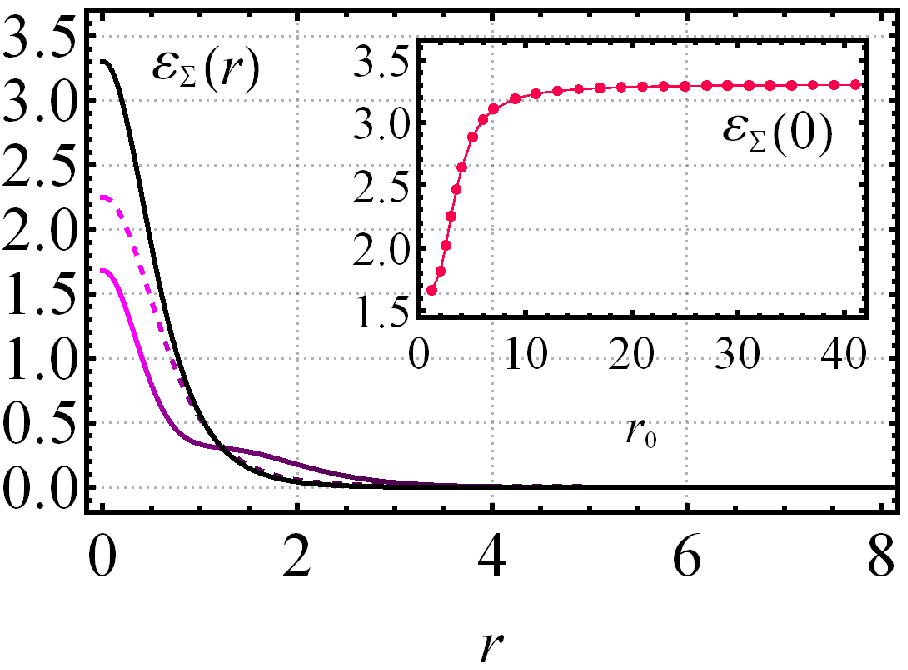}\hspace{0.13cm} %
	\includegraphics[width=4.2cm]{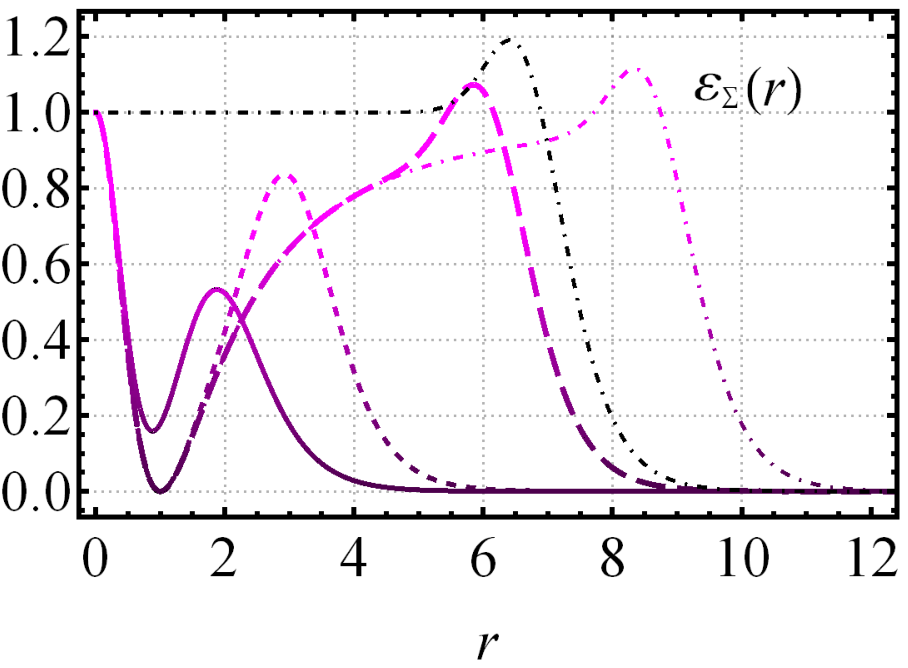} \vspace{-0.6cm}
	\caption{The profiles for the energy density $\varepsilon_{_\Sigma} (r)$ in  both cases in the presence (color line), by fixing $\alpha=1$, and absence (black line) of the dielectric function (\ref{27B}). On the left, it is depicted $N=1$, $r_{0}=1$ (solid line) and $r_{0}=3$ (dashed line); 		(inset) we have $\varepsilon_{_{\Sigma}}(0)$ vs. $r_{0}$ for the dielectric function (\ref{27B}) with $N=1$. On the right, profiles for $r_{0}=1$ with $N=2$ (solid line), $N=5$ (dashed line), $N=20$ (long-dashed line), and $N=30$
(dot-dashed line).}
	\label{Fig08}
\end{figure}

Figure \ref{Fig08} depicts the profiles of the energy density $\varepsilon_{_\Sigma}$ that allows us to analyze, in the present scenario, its main features. Our numerical results are better understood or complemented by the behavior at the boundary values. {Thus, near the origin}, it is given by
\begin{eqnarray}
\varepsilon_{_\Sigma}(r) &\approx&1+N^{2}(f_{N}) ^{2}r^{2N-2} -\frac{(N^{2} +2N+2) (f_{N}) ^{2}r^{2N}}{2} \;\; \notag\\[0.2cm]
&&- \frac{4 r^{2\alpha }}{r_{0}^{2\alpha }} -\frac{N^{2}(f_{N})^{4}r^{4N-2}}{2} +\frac{16\alpha ^{2}r^{4\alpha -2}}{r_{0}^{4\alpha }},
\end{eqnarray}
guaranteeing at least the two first lowest-order terms.  Further, for $r\rightarrow\infty$, $\varepsilon_{_\Sigma}$ behaves as
\begin{equation}
\varepsilon_{_\Sigma}(r) \approx 2(\mathcal{C}_{\infty})^2 r^{-1}e^{-2r}.
\label{ebpsII}
\end{equation}

On the left-hand side of Fig. \ref{Fig08}, for $N=1$, the profiles of $\varepsilon_{\Sigma}$ are lumps centered at the origin whose amplitude  $\varepsilon_{\Sigma}(0)=1+(f_1)^2$ increases as $r_{0}$ grows until that, for sufficiently large values of $r_{0}$ (see inset),  it saturates attaining the value corresponding to the standard sigma model (the solid black line). Conversely, for $N\geq 2$, we will always have $\varepsilon_{\Sigma}(0) =1$, as shown on the right of Fig. \ref{Fig08}. There, for a fixed $r_0$ and increasing values of $N$, we observe the profiles of $\varepsilon_{\Sigma}$ acquire a local minimum at $r^*<r_{0}$ that becomes null and remains localized at $r*=r_0$ for all $N\geq N^{*}$ (e.g., $N^{*}=5$ in Fig. \ref{Fig08}).


\subsection{Third scenario\label{Fs3}}

Motivated by Ref. \cite{B_Research_2019}, we now consider the dielectric function as follows
\begin{equation}
\Sigma (\chi) =\frac{1+\lambda ^{2}}{\lambda ^{2}+\cos
^{2}\left( m\pi \chi \right) }\text{,}  \label{28B}
\end{equation}%
where $m\in\mathbb{N}$ and $\lambda \in\mathbb{R}$.  We remark that for very large values of $\lambda $ the dielectric function $\Sigma (\chi) \rightarrow 1$ which means we recover the  standard sigma model.

As in the previous scenarios, we have a set of BPS equations,
\begin{eqnarray}
f^{\prime }&=&\frac{a}{r}\sin f\text{,}
\end{eqnarray}
\begin{eqnarray}
-\frac{a^{\prime }}{r}&=&\displaystyle\frac{\lambda ^{2}+\cos^{2} \left( \displaystyle m\pi \frac{r^{2\alpha } -r_{0}^{2\alpha }}{r^{2\alpha }+r_{0}^{2\alpha }}\right)}{1+\lambda ^{2}}  \cos f,
\end{eqnarray}
whose solutions, under the boundary conditions (\ref{4B2}) and (\ref{4B3}), provide the self-dual solitons in the current scenario.

{We now compute the approximated solutions for the field profiles, which characterize their behaviors near the boundaries.  At the origin, the sigma field and gauge field profiles behave as
\begin{eqnarray}
f(r)&\approx& f_{N}r^{N}-\frac{f_{N}r^{N+2}}{4}-\frac{(f_{N})^{3}r^{3N}} {12}\nonumber\\[0.2cm]
& & +\ldots +\frac{m^{2}\pi ^{2}f_{N}r^{N+4\alpha+2}}{(2\alpha+1)^2( \lambda ^{2}+1) r_{0}^{4\alpha}}, \quad\quad\\[0.3cm]
a(r) &\approx&N-\frac{r^{2}}{2}+\frac{(f_{N})^{2}r^{2N+2}}{4( N+1) } \nonumber\\[0.2cm]
&&+...+\frac{2 m^{2}\pi ^{2}r^{4\alpha +2}}{(2\alpha+1)( \lambda ^{2}+1)
r_{0}^{4\alpha }},
\end{eqnarray}
respectively, where $f_{N}$ is a positive real number. The expressions above guarantee at least the two lowest-order terms and the lowest contribution coming from the dielectric function. On the other hand, the behaviors for $r\rightarrow\infty$  coinciding with the ones obtained in Eqs. (\ref{case2inf_1}) and (\ref{case2inf_2}), respectively.

\begin{figure}[t]
	\includegraphics[width=4.2cm]{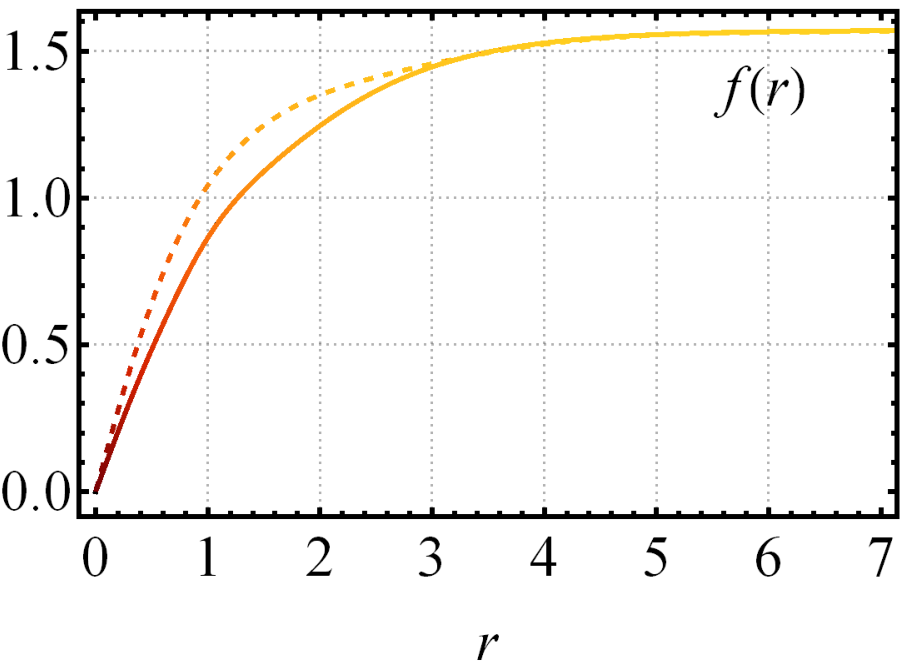}\hspace{0.14cm} %
	\includegraphics[width=4.2cm]{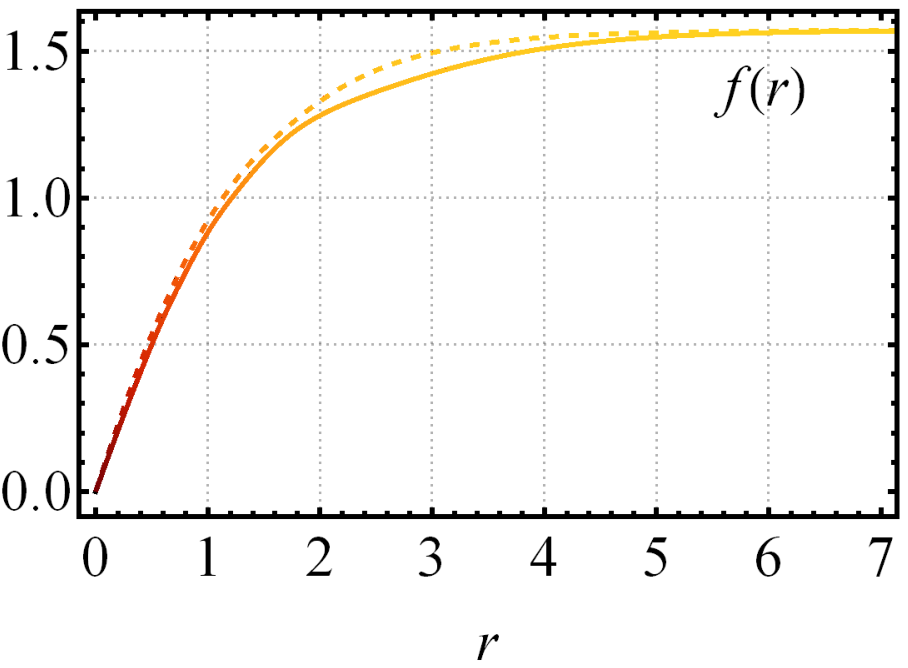} \vspace{-0.6cm}
	\caption{The sigma field {profiles} $f (r)$ for $N=1$ and $\lambda=0$. (left) Profiles for $\alpha=1$, $m=1$ with $r_{0}=1$ (solid line) and $r_{0}=5$ (dashed line). (right) Profiles for $%
		r_{0}=1$, $m=2$ with $\alpha=1$ (solid line) and $\alpha=2$
		(dashed line).}
	\label{Fig09}
\end{figure}

For the magnetic field and the energy density $\varepsilon_{_ \Sigma} (r)$, the behaviors in $r=0$ obey the expressions given by
\begin{eqnarray}
B(r)&\approx&  1-\frac{(f_{N})^{2}}{2} r^{2N} +...+\frac{4 m^{2}\pi ^{2}r^{4\alpha}}{(\lambda ^{2}+1) r_{0}^{4\alpha }}, \\[0.3cm]
\varepsilon _{_{\Sigma }}(r) &\approx&1+N^{2}(f_{N})^{2}r^{2N-2}-%
\frac{(N^{2}+2N+2)(f_{N})^{2}r^{2N}}{2} \;\notag \\[0.2cm]
&&-\frac{N^{2}(f_{N})^{4}r^{4N-2}}{2}+...-\frac{4m^{2}\pi
^{2}r^{4\alpha }}{(\lambda^{2}+1) r_{0}^{4\alpha }},
\end{eqnarray}
respectively. Already, for $r\rightarrow\infty$,  the behaviors are the very same as the ones given in Eqs. (\ref{BiiInfy}) and (\ref{ebpsII}), respectively, found in the previous scenario.}

Next, we investigate numerically how the dielectric function (\ref{28B}) modifies the soliton profiles of the standard sigma model.  The analysis considers two situations depending on the $\lambda$ parameter: the first one associated with  $\lambda =0$ and the second case related to $\lambda \neq 0$.

\begin{figure}[t]
	\includegraphics[width=4.2cm]{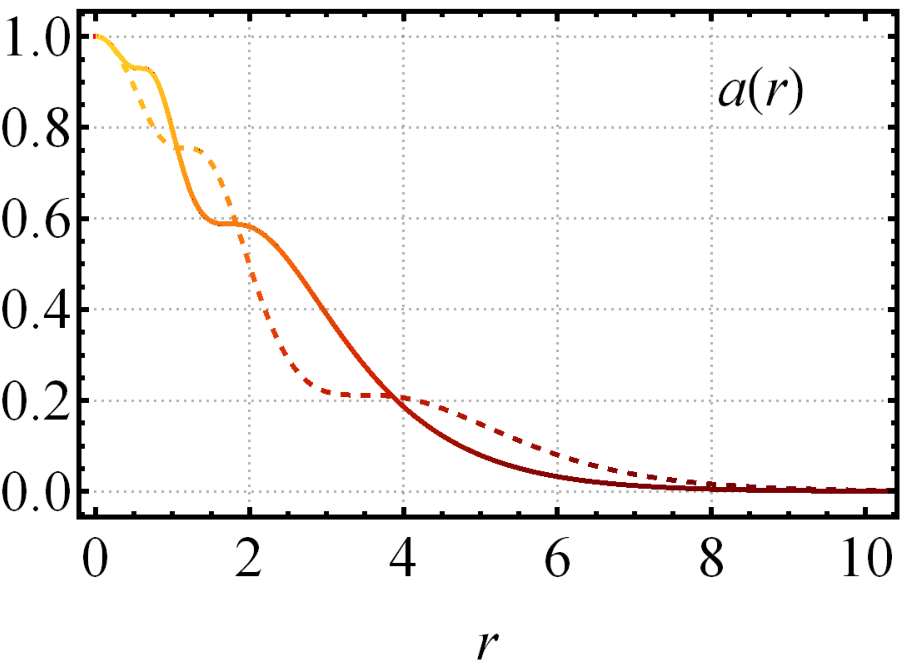}\hspace{0.14cm} %
	\includegraphics[width=4.2cm]{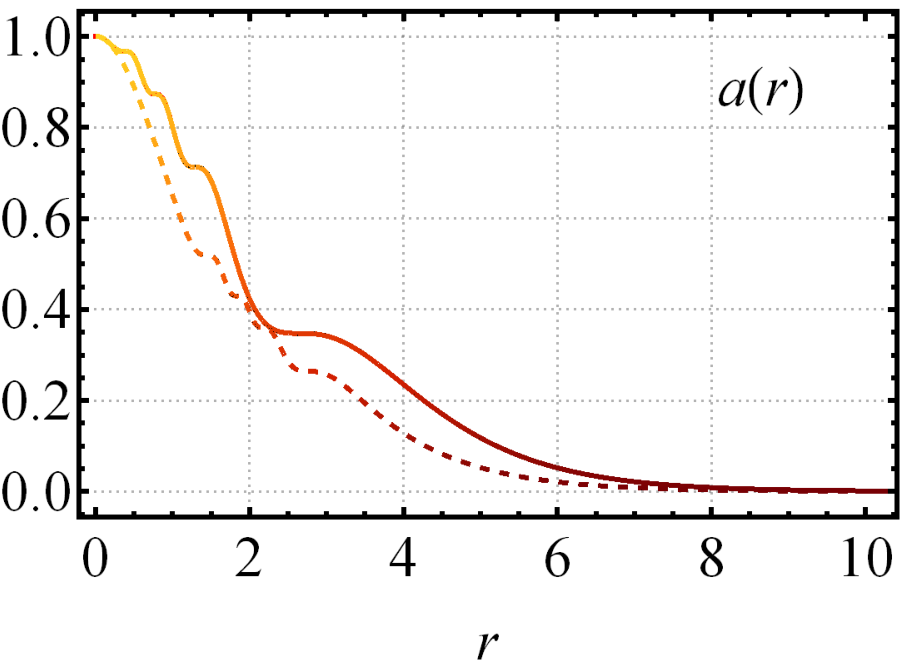} \vspace{-0.6cm}
	\caption{The gauge field profiles $a (r)$ with fixed values $N=1$ and $%
		\lambda=0$. (left) Profiles for $\alpha=1$, $m=1$ with $%
		r_{0}=1$ (solid line) and $r_{0}=2 $ (dashed line). (right) Profiles for $%
		r_{0}=1$, $m=2$ with $\alpha=1$ (solid line) and $\alpha=3$
		(dashed line).}
	\label{Fig10}
\end{figure}

\begin{figure}[b]
	\includegraphics[width=4.2cm]{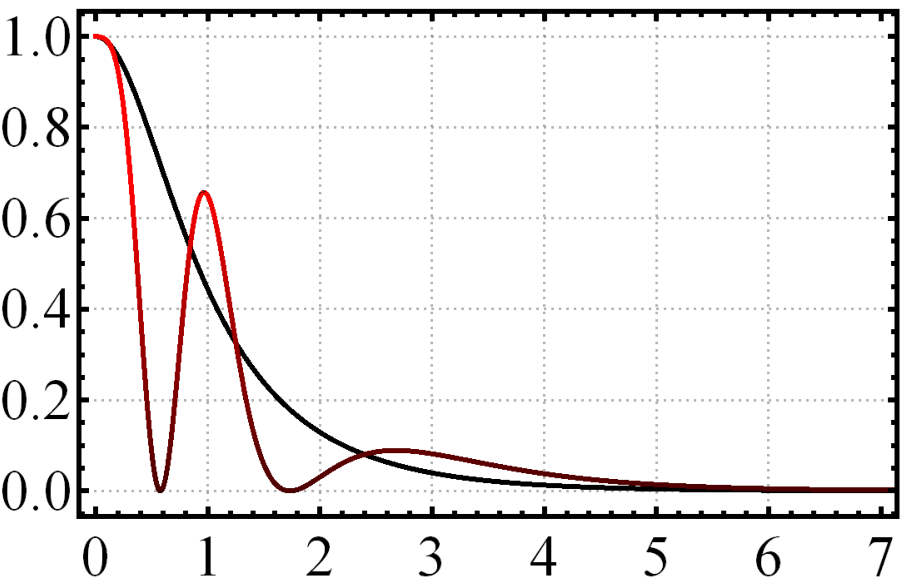}\hspace{0.14cm} %
	\includegraphics[width=4.2cm]{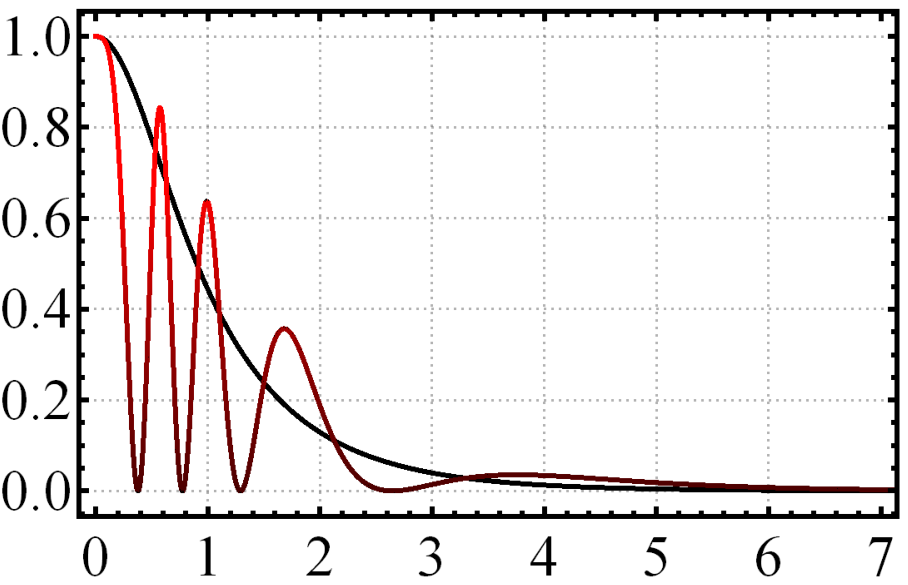}\vspace{0.2cm} %
	\includegraphics[width=4.2cm]{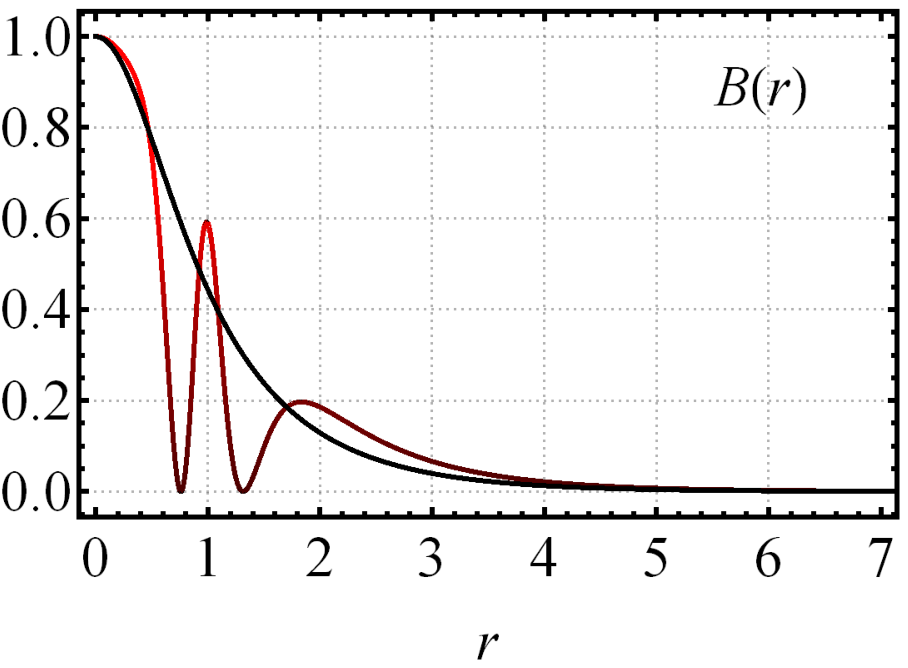}\hspace{0.14cm} %
	\includegraphics[width=4.2cm]{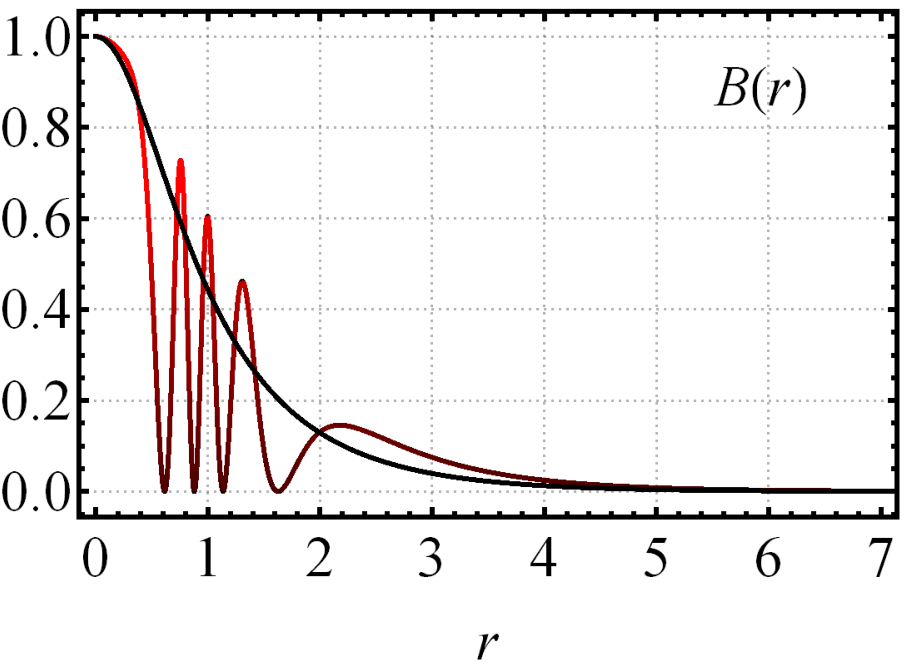}\vspace{-0.1cm}
	\caption{The magnetic field profiles for both in the presence (color line) and absence (black line) of the dielectric function (\ref{28B}). We have $\lambda=0$, $N=1$, $r_{0}=1$, $\alpha=1$ (top) and $\alpha =2$ (bottom) with $m=1$ (left) and $m=2$ (right).}
	\label{Fig11}
\end{figure}

\begin{figure}[t]
	\includegraphics[width=3.1cm]{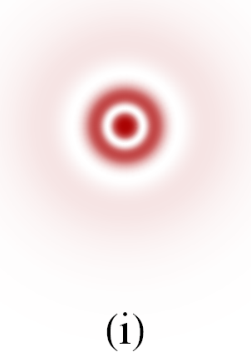}\hspace{-0.35cm} %
	\includegraphics[width=3.1cm]{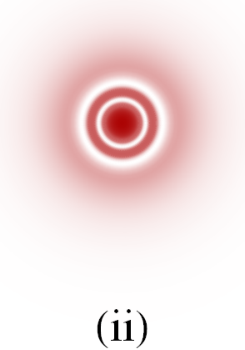}\hspace{-0.55cm} %
	\includegraphics[width=3.1cm]{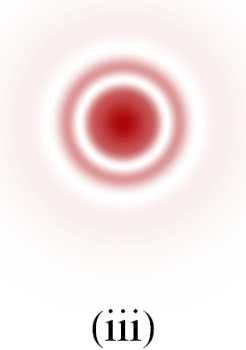}\hspace{-0.35cm} %
	\includegraphics[width=3.1cm]{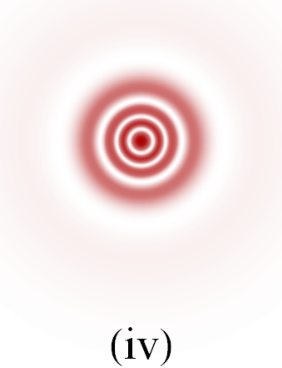}\hspace{-0.39cm} %
	\includegraphics[width=3.1cm]{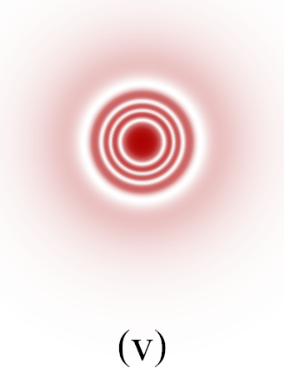}\hspace{-0.458cm} %
	\includegraphics[width=3.1cm]{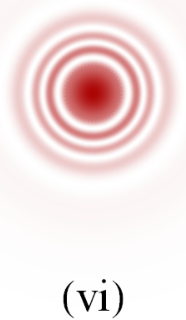}\vspace{0.cm}
	\caption{The magnetic field depicted in the plane with values fixed for $N=1$
		and $\lambda=0$. We display $m=1$, $r_{0}=1$, $\alpha=1$ (i)
		and $\alpha=2$ (ii), and $r_{0}=2$, $\alpha=2$ (iii).
		Following, we have $m=2$, $r_{0}=1$, $\alpha=1$ (iv) and $%
		\alpha=2$ (v), and $r_{0}=2$, $\alpha=2$ (vi).}
	\label{Fig12}
\end{figure}
\begin{figure}[b]
	\includegraphics[width=4.2cm]{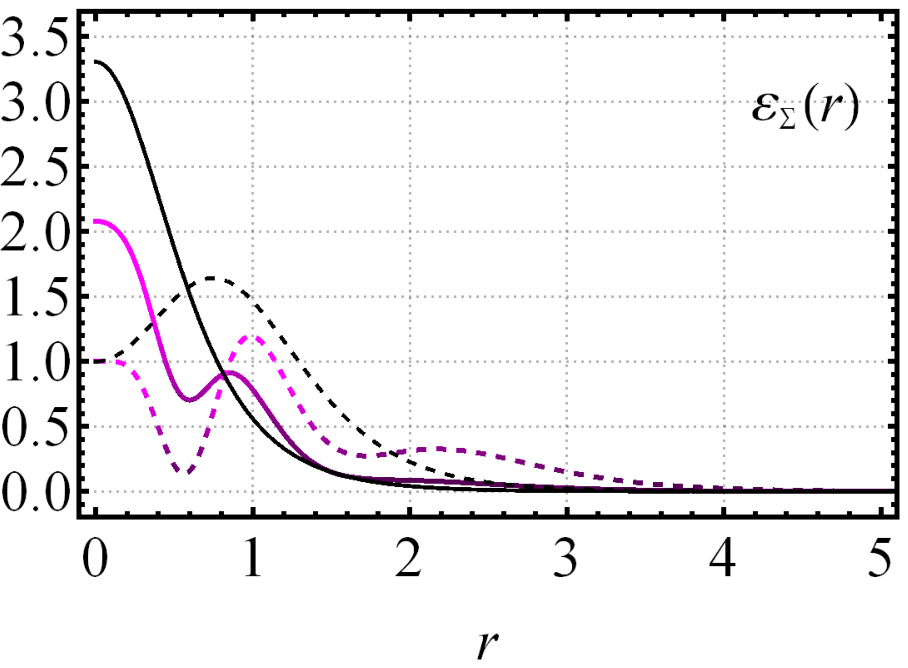}\hspace{0.14cm} %
	\includegraphics[width=4.2cm]{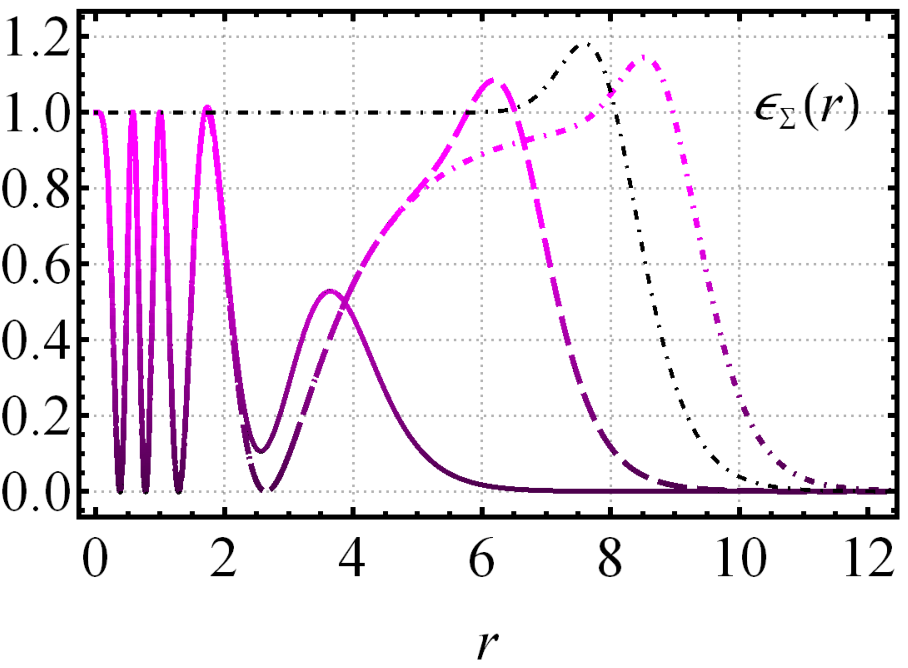} \vspace{-0.6cm}
	\caption{The profiles for energy density $\varepsilon_{_\Sigma} (r)$ in both cases in the presence (color line) and absence (black line) of the dielectric function (\ref{28B}), by setting $\lambda=0$, $\alpha=1 $, $r_{0}=1$. On the left, profiles for $m=1$, $N=1 $ (solid line) and $N=2$ (dashed line). On the right, for $m=2$, $N=5$ (solid line), $N=20$ (long-dashed line) and $N=40$ (dot-dashed line).}
	\label{Fig13E}
\end{figure}

\begin{figure}[t]
	\includegraphics[width=4.2cm]{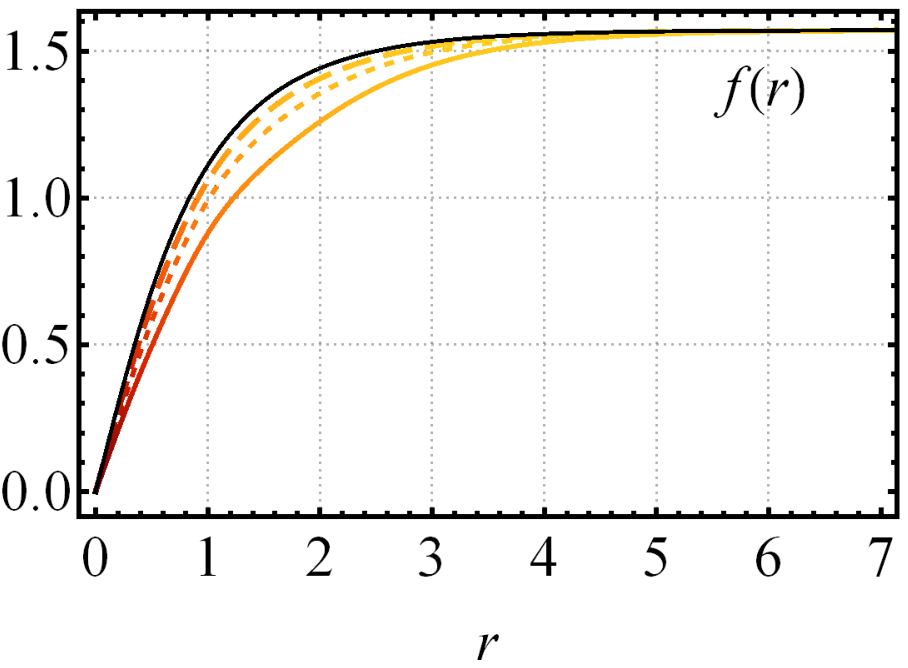}\hspace{0.14cm} %
	\includegraphics[width=4.2cm]{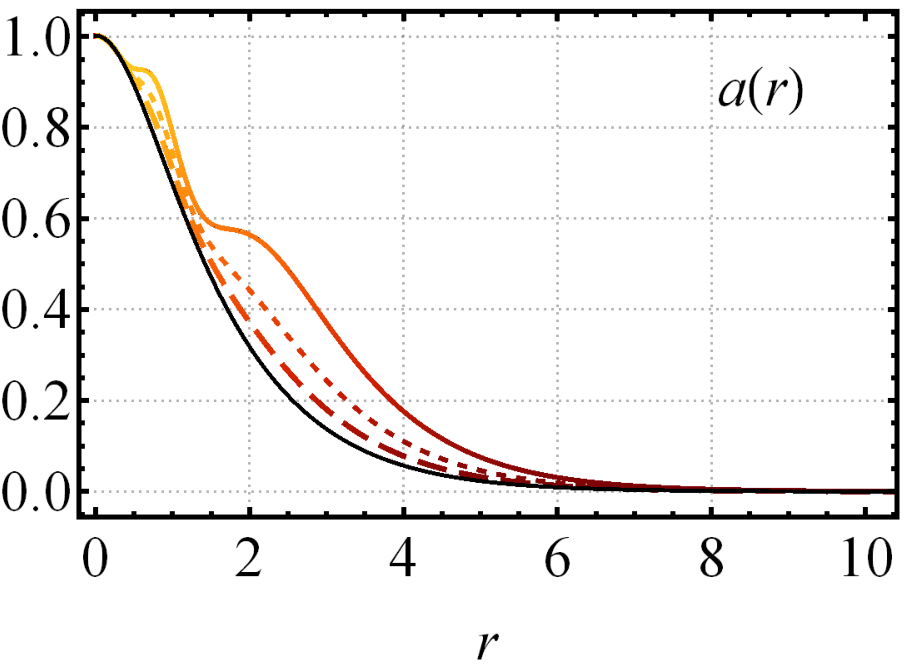} \vspace{-0.6cm}
	\caption{The solitons solution $f(r)$ (left) and $a\left(r\right)$ (right), comparing with the standard sigma model (solid black line), all with $N=1$. Under the dielectric function (\ref{28B}), we set $\alpha=1$, $r_{0}=1$, $m=1$ for $\lambda=0.2$ (color solid line), $\lambda=0.8$ (dashed line) and $\lambda=1.5$ (long-dashed line).}
	\label{Fig14}
\end{figure}
\subsubsection{Simplest case: $\lambda=0$}

In the current scenario, the Figs. \ref{Fig09} and \ref{Fig10} depict the numerical profiles for the sigma and gauge fields. We highlight the role played by the parameters $m$ and $\alpha $ whose effects are more notorious in the gauge field profiles. The first one causes the arising of $2m$-plateaus and the second one profiles more localized around the origin when $\alpha$ grows, respectively, both producing relevant effects in the magnetic field structure.  In this sense,  Figs. \ref{Fig11} and \ref{Fig12} show the magnetic field profiles for distinct values of the parameters $r_{0}$, $\alpha$, and $m$, comparing with the corresponding profile of the standard sigma model (black lines). Similarly to the previous scenarios, we notice the parameter $r_{0}$ controls both the intensity and width of the structures, while the radius of their respective cores increases as $\alpha$ grows. Nevertheless,  the additional parameter $m$ associated with the $2m$-plateaus of the gauge field determines an equal number of external rings to the core.

The effects of $r_0$ and $\alpha$ on the energy density $\varepsilon_{_{ \Sigma}}$ are similar, albeit less intense, to the ones exhibited by the magnetic field in Fig. \ref{Fig11}. This way, in  Fig. \ref{Fig13E}, we only draw the profiles for different values of  $N$ and $m$, and some corresponding to the ones in the standard sigma model (black lines). At the origin, the density $\varepsilon_{_{\Sigma}}$ is always {nonnull} and has the following behavior according the $N$ values: it is  $\varepsilon_{_{\Sigma}}(0)= 1+(f_{1})^{2}$ for $N=1$ and $\varepsilon _{_{\Sigma }}(0)=1$ for $N\geq 2$. On the right-hand side of the figure, such as in the magnetic field case, we see several ring-like structures that are associated with the $m$ value and become more noticeable when $N\geq 2$. We also observe that for large values of $N$, the border of the profile follows the format of the one belonging to the standard sigma model  (on the right, e.g., see the profile for $N=40$).

\begin{figure}[b]
	\includegraphics[width=3.5cm]{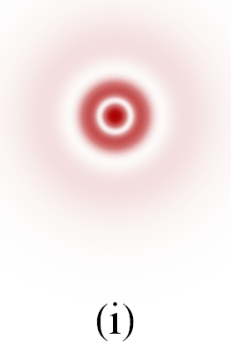}\hspace{0.cm} %
	\includegraphics[width=3.5cm]{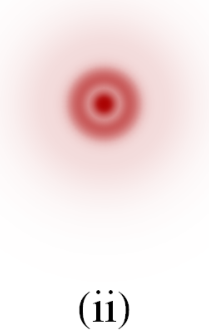}\hspace{0.cm} %
	\includegraphics[width=3.5cm]{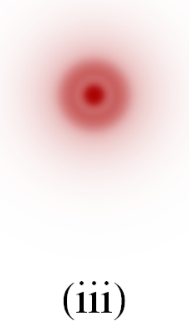}\vspace{0.cm}
	\includegraphics[width=3.5cm]{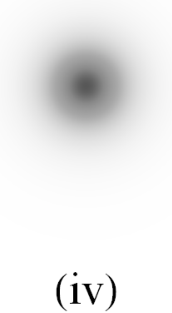}\vspace{0.cm}
	\caption{The magnetic field depicted in the plane for both in the presence (color plot) and absence (black plot) of the dielectric function (\ref{28B}). The conventions are as in Fig. \ref{Fig14}, being $\lambda=0.2$ (i), $\lambda=0.8$ (ii) $\lambda=1.5$ (iii) and standard sigma model (iv). }
	\label{Fig15}
\end{figure}

\subsubsection{Nonvanishing case: $\lambda \neq 0$}

We now present a brief analysis of the case $\lambda\neq0$. As already mentioned, for sufficiently large values of $\lambda$, we recuperate the standard gauged sigma model, i.e., the dielectric function approaches to the unit value. Fig. \ref{Fig14} shows such behavior for the profiles of the sigma and gauge fields, which become more localized around the origin as $\lambda$ increases. In Fig. \ref{Fig15}, the planar depict of the magnetic field also shows such a diminishing effect. Consequently, in the present scenario, the internal structure diminishes as $\lambda$ increases until that, in the limit of sufficiently large values, the configurations match the ones of the standard gauged sigma model.

\section{Conclusions and remarks}

\label{conclusion}

We have studied the existence of new topological magnetic solitons living into a dielectric medium $\Sigma$. To describe these structures, we have considered a gauged sigma model containing an extra real scalar field (source field $\chi$) that characterizes the dielectric properties of the medium, $\Sigma\equiv\Sigma(\chi)$. {Stable solitons are obtained by following the BPS technique,} whose implementation is only possible by introducing a superpotential $\cal{W}(\chi)$ in the source sector, allowing us to decouple it from both the gauge and sigma sectors.

Even without knowing an explicit form of both the dielectric function $\Sigma(\chi)$ and superpotential $\mathcal{W}(\chi)$, we have found first-order equations, which  are also solutions of the Euler-Lagrange ones. Because of the arbitrariness of these functions, we point out that the proposed model can fit a variety of physical environments. In our analysis, we have adopted the superpotential (\ref{24B}) that supports kink-like solutions for the source field that allows us to investigate solitons with internal structure in three distinct scenarios.   The superpotential introduces the parameter $\alpha>0$ and, by solving the BPS equation (\ref{21B}), we obtain the kink solution (\ref{25B}) presenting a second parameter $r_{0}$, which characterizes the kink radius where $\chi(r_0)=0$. Further, the dielectric function $\Sigma(\chi)$, theoretically, could introduce a set of additional parameters.

{In the first scenario, we only study the case $\alpha=1$ because we have perceived there are not solutions valid for $\alpha\geq 2$. Here, the kink } radius $r_{0}$ indicates the position of the maximum amplitude attained by both the magnetic field and energy density $\varepsilon_{_\Sigma}$ for sufficiently large winding numbers. Additionally, we found that the profiles are quite similar to the ones of the Chern-Simons vortices \cite{Jackiw_1990}, the ones in some generalized Maxwell-Higgs model \cite{Maxwell_Higgs_SE}, and the solitons with an internal structure of the gauged $CP(2)$ model studied in Ref. \cite{Andrade_2019}.

In the second scenario, we have discussed the formation of solitons whose magnetic field has a distribution composed by a core with a maximum amplitude around which is formed a ring with a maximum having a lower intensity.  The magnetic field is null at $r=r_{0}$, so it separates the core from the ring. Furthermore, {the magnetic field profiles} in the range $0\leq r \leq r_{0}$ remind the ones found by Nielsen-Olesen \cite{Nielsen_1973}, whereas for $r>r_0$  resemble the behavior of the Chern-Simons vortices, including typical exponential decay, here occurring yet in the presence of the dielectric medium. Also, we have realized the parameter $\alpha$ plays the role of controlling the core size and the maximum intensity of the external ring.   On the other hand, the profiles of the energy density $\varepsilon_{_\Sigma}$, for sufficiently large values of $N$, also vanish at $r_0$ and present a behavior very similar to the one described for the magnetic field.

{In the third and last scenario,}  we have introduced an oscillating dielectric function controlled via two extra parameters ($m\in \mathbb{Z}$ and $\lambda\in \mathbb{R}$) to describe a multilayered system. The value $\lambda=0$ generates a dielectric system producing a more notable formation of the internal structures than for $\lambda\neq 0$. In the latter case, the emergent internal structure becomes more relevant for small values of $\lambda$ than {the big ones, such that, for sufficiently large values, the} dielectric effects diminish rapidly and, so we recover the realm of the usual gauged sigma model. In both cases, {the arising of} $2m$-plateaus {along the gauge} field profile implies in the same number of magnetic rings {surrounding the soliton-core.}

Finally, we are studying the possible existence of BPS charged vortices (maybe behaving as anions) in the presence of dielectric media in the realm of the gauged $O( 3)$ sigma model with Chern-Simons term \cite{Ghosh_1996}. Advances in this direction will report elsewhere.

\begin{acknowledgments}
This study was financed in part by the Coordena\c{c}\~ao de Aperfei\c{c}oamento de Pessoal de N\'{\i}vel Superior - Brasil (CAPES) - Finance Code 001. We thank also the Conselho Nacional de Desenvolvimento Cient{\'\i}fico e Tecnol\'ogico (CNPq), and the Funda\c{c}\~ao de Amparo \`a Pesquisa e ao Desenvolvimento Cient{\'\i}fico e Tecnol\'ogico do Maranh\~ao (FAPEMA) (Brazilian Government agencies). R. C.  acknowledges the support from the grants CNPq/306724/2019-7, CNPq/423862/2018-9 and FAPEMA/Universal-01131/17. In particular, A. C. S. and M. L. D. thank the full support from CAPES.
\end{acknowledgments}

\end{document}